\documentclass[useAMS,usenatbib]{mn2e}
\usepackage{amssymb,epsfig,natbib}
\title[SN 2004A: Another Type II-P Supernova with a Red Supergiant Progenitor]
{SN 2004A: Another Type II-P Supernova with a Red Supergiant Progenitor}
\author[M. A. Hendry {\rm et al.}]
{M. A. Hendry$^1$, S. J. Smartt$^{2}$, R. M. Crockett$^2$, J. R. Maund$^3$, A. Gal-Yam$^{4,5}$,
\newauthor D.-S. Moon$^6$, S. B. Cenko$^6$, D. W. Fox$^4$, R. P. Kudritzki$^7$, C. R. Benn $^8$, 
\newauthor R. \O stensen$^8$ \\
\\
$^1$Institute of Astronomy, University of Cambridge, Madingley Road,
        Cambridge CB3 0HA\\
$^2$Department of Physics and Astronomy, Queen's University Belfast, Belfast BT7 1NN\\
$^3$Astronomy Department, University of Texas, 1 University Station, C1400, Austin, TX 78712\\
$^4$Department of Astronomy, MS 105-24, California Institute of Technology, Pasadena, CA 91125, USA\\
$^5$Hubble Fellow\\
$^6$Department of Physics and Space Radiation Laboratory, MS 220-47, California Institute of Technology, Pasadena, CA 91125, USA\\
$^7$Institute for Astronomy, University of Hawaii, 2680 Woodlawn Drive, Honolulu, Hawaii 96822, USA\\
$^8$Isaac Newton Group, Apartado 321, E-38700 Santa Cruz de La Palma, Espa\~{n}a\\
}
\date{Submitted 24 Nov 2005}

\pubyear{2005}

\def\LaTeX{L\kern-.36em\raise.3ex\hbox{a}\kern-.15em
  T\kern-.1667em\lower.7ex\hbox{E}\kern-.125emX}

\def \aj {AJ}
\def \mnras {MNRAS}
\def \apj {ApJ}
\def \apjl {ApJL}
\def \aap {A\&A}
\def \nat {Nature}

\def \iauc {IAUC}
\def \pasp {PASP}

\def \apjs {ApJS}
\def \aaps {A\&AS}

\bibpunct{(}{)}{;}{a}{}{,}

\newcommand{\msun}{\mbox{M$_{\odot}$}}
\newcommand{\hub}{\mbox{$\rm{km}\,s^{-1} {\rm Mpc}^{-1}$}}
\newcommand{\kms}{\mbox{$\rm{km}\,s^{-1}$}}

\newcommand{\ms}{\mbox{$M_{\rm MS}$}}
\newcommand{\mni}{\mbox{$M_{\rm Ni}$}}
\newcommand{\mej}{\mbox{$M_{\rm ej}$}}
\newcommand{\teff}{\mbox{$T_{\rm eff}$}}
\newcommand{\ebv}{\mbox{$E(B\!-\!V)$}}
\newcommand{\bv}{\mbox{$B\!-\!V$}}
\newcommand{\vr}{\mbox{$V\!-\!R$}}
\newcommand{\vi}{\mbox{$V\!-\!I$}}

\begin{document}

\label{firstpage}

\maketitle

\begin{abstract}
We present a monitoring study of SN~2004A and probable discovery of a progenitor star in pre-explosion {\it HST} images. The photometric and spectroscopic monitoring of SN~2004A show that it was a normal Type II-P which was discovered in NGC~6207 about two weeks after explosion. We compare SN~2004A to the similar Type II-P SN~1999em and estimate an explosion epoch of 2004~January~6. We also calculate three new distances to NGC~6207 of $21.0\pm4.3$, $21.4\pm3.5$ and $25.1\pm1.7$\,Mpc. The former was calculated using
the Standard Candle Method (SCM) for \mbox{SNe II-P}, and the latter two from the 
Brightest Supergiants Method (BSM). We combine these three distances
with existing kinematic distances, to derive a mean value of \mbox{$20.3\pm
3.4$\,Mpc}. Using this distance we estimate that the ejected nickel mass 
in the explosion is $0.046^{+0.031}_{-0.017}$\,\msun. 
The progenitor of SN~2004A is identified in pre-explosion WFPC2 
F814W images with a magnitude of $m_{\rm F814W}=24.3\pm0.3$, but 
is below the detection limit of the F606W images. We show that 
this was likely a red supergiant (RSG) with a mass of $9^{+3}_{-2}$\,\msun. 
The object is detected at 4.7$\sigma$ above the background noise. 
Even if this detection is spurious, the 5$\sigma$ upper limit would give
a robust upper mass limit of 12\,\msun\ for a RSG progenitor. 
These initial masses are very similar to those of two previously 
identified RSG progenitors of the Type II-P SNe
2004gd ($8^{+4}_{-2}$\,\msun) and 2005cs ($9^{+3}_{-2}$\,\msun). 
\end{abstract}


\section{Introduction}

Supernovae (SNe) are associated with the deaths of stars, in
particular core-collapse supernovae (CCSNe) are associated with the
deaths of massive stars, which have initial masses greater than about
8\,\msun. 
SNe are principally separated into two categories, those without hydrogen
(Type I) and those with (Type II). Only Type Ia SNe are thought to be
thermonuclear explosions, which arise from accreting white dwarfs in
binary stellar systems. 
All the other sub-types are thought to
be initiated by the core collapsing in massive stars. The type of SN
that occurs depends on the massive star's evolutionary stage at the
time of the explosion. The plateau subclass of Type II SNe (SNe II-P)
are thought to arise from the explosions of red supergiants (RSGs),
which have initial masses greater than 8--10\,\msun\ and have retained their
hydrogen envelopes before core collapse 
\citep{1976ApJ...207..872C, 1985PhRvL..55..126B, 2003ApJ...591..288H,2004MNRAS.353...87E}. 

Until the discovery of the red supergiant (RSG) that exploded as
SN~2003gd \citep{2003PASP..115.1289V,2004Sci...303..499S,2005MNRAS.359..906H}, there had
been no direct confirmation that SNe II-P did indeed arise from the
explosions of RSGs. Before this detection there had been only two other
unambiguous detections of Type II progenitors, neither of which fitted
the evolutionary scenario that is commonly accepted. These were the
progenitors of the peculiar Type II-P SN~1987A, which was a blue supergiant \citep[BSG,][]{1987Natur.327...36W,1989A&A...219..229W}, and
the Type IIb SN~1993J that arose in a massive interacting binary
system
\citep{1994AJ....107..662A,2002PASP..114.1322V,2004Natur.427..129M}. The recent
discovery of SN~2005cs (II-P) in M51, a galaxy with 
deep multi-colour pre-explosion images from the {\it Hubble Space Telescope} ({\it HST}), led to the discovery of another RSG progenitor of a II-P 
\citep{2005MNRAS.tmpL..88M,2005astro.ph..7394L}. The estimated mass of the 
star was $\ms=9^{+3}_{-2}$\,\msun, similar to the 
mass ($\ms=8^{+4}_{-2}$\,\msun) for the progenitor  of SN~2003gd 
\citep{2004Sci...303..499S, 2003PASP..115.1289V}. 
A supergiant of mass 
$\ms=15^{+5}_{-2}$\,\msun\ was found to be coincident with the 
Type II-P SN~2004et by \citet{2005PASP..117..121L}, although it is likely
not to have been as cool as an M-type supergiant, and the SN itself
may be peculiar. There have been other extensive attempts to detect 
progenitors of nearby SNe on ground- and space-based archival
images e.g.  
\citet{2005MNRAS.360..288M,2003MNRAS.343..735S,2003PASP..115....1V,2004ApJ...615L.113M}, 
which have set upper mass limits mostly on II-P events. The low mass of the 
progenitors discovered and upper limits set has led to the suggestion that 
SNe II-P come only from RSGs with masses less than about 15\,\msun\
\citep{jrmthesis,2005astro.ph..7394L}

SN~2004A is another example of a nearby SN II-P which 
has {\it HST} pre-explosion images, allowing the search for a progenitor star. 
SN~2004A was discovered by K. Itagaki of Teppo-cho, Yamagata, Japan on
January 9.84 {\sc ut} using a 0.28-m f/10 reflector. Itagaki confirmed his
discovery on January 10.75 {\sc ut}, with a location of R. A. $=
16^h43^m01.90^s$, Dec. $= +36^\circ50\arcmin12.5\arcsec$, around 22
arcsec west and 17 arcsec north of the centre of NGC~6207. Itagaki
reported that no object was visible on his observations of 2003
December~27, which had a limiting magnitude of 18, or any of his
observations prior to this date \citep{2004IAUC.8265....1N}. Itakagi's
observations allow the explosion epoch to be fairly well constrained,
suggesting that SN 2004A was discovered when it was quite young at
less than 14\,d after explosion. An optical spectrum was obtained by
\citet{2004IAUC.8266....2K} on January 11.8 and 11.9 {\sc ut}, and showed a
blue continuum with P-Cygni profiles of the Balmer lines, consistent
with a Type II SN. The emission features were somewhat weak suggesting
that the SN was indeed young, in line with Itagaki's
observations. The expansion velocity, measured from the minima of the
Balmer lines, was around 12\,000\,\kms. In {\it HST} Cycle 10, we had a snapshot 
programme to enhance the {\it HST} archive with 100--200 Wide Field Planetary Camera 2 (WFPC2) multi-colour images of galaxies within approximately 20\,Mpc. In the future, SNe discovered in these galaxies could have pre-explosion images available to constrain the nature of the progenitor stars. This strategy is now beginning to bear fruit, NGC~6207 was one of those targets and the pre-explosion site of SN~2004A was imaged in three filters.  

In this paper we present photometric and spectroscopic data of
SN~2004A in Section~\ref{sec:obs} followed by an analysis of the
photometry in Section~\ref{sec:epoch}, where an explosion date is
estimated. We estimate the reddening towards the SN in
Section~\ref{sec:red} and obtain the expansion velocity in
Section~\ref{sec:expvel}. The distance to NGC~6207 is not well known
and only two distance estimates, which are both kinematic, exist in
the literature. We estimate the distance using two further methods
and compile the distances within the literature, in an attempt to
improve the situation, in Section \ref{sec:D}. Using the distance
found we then calculate the amount of nickel synthesised in the
explosion in Section~\ref{sec:Ni}. We present the discovery of the
progenitor in Section~\ref{sec:prog} and a discussion of the implications
and conclusion in Sections~\ref{sec:diss} and \ref{sec:con},
respectively. Throughout this work we have assumed the galactic
reddening laws of \citet*{1989ApJ...345..245C} with $R_V = 3.1$.

\section{Observations}\label{sec:obs}

\subsection{Ground-based photometry of SN~2004A}\label{sec:phot}

{\it BVRI} photometry was obtained shortly after discovery from the
following telescopes: the 2.0-m Liverpool Telescope (LT), La Palma; 
the 4.2-m William Herschel Telescope (WHT), La Palma; and 
the Robotic Palomar 60-inch telescope 
(P60\footnote{http://www.astro.caltech.edu/$\sim$derekfox/P60/}; 
Cenko et al. 2006, in prep.) as part of the Caltech Core-Collapse Program\footnote{http://www.astro.caltech.edu/$\sim$avishay/cccp.html}
\citep[CCCP,][Gal-Yam et al. 2006 in prep.]{2004AAS...205.4006G,2005ApJ...630L..29G}
The LT
observations were taken with the optical CCD Camera, RATCam, using its
Bessel {\it BV} and Sloan $r\arcmin i\arcmin$ filters. The data were
reduced using the LT data reduction pipeline. The WHT observations were taken with the
Auxiliary Port Imaging Camera (AUX), using its {\it BVRI} filters, and
were reduced using standard techniques within {\sc iraf}. 
The frames
were debiased and flat-fielded using dome flats from a few nights
later. Details of the P60 camera and data reduction can be found in 
\citet{2005PASP..117..132R}. 
A summary of these observations can be found in Table
\ref{tab:phot} as well as the results from the SN photometry.

\begin{table*}
  \caption[]{Journal and results of optical photometry of SN~2004A.}
  \begin{minipage}{.9\linewidth}
    \begin{center}
      \begin{tabular}{lrrrrrrl} \hline
        Date & JD &  Phase & $B$ & $V$ & $R$ & $I$ & Telescope + \\
        & (245\,0000+) & (days) & & & & &Instrument \\
        \hline
2004 Feb 01 & 3036.77 &  26 & 16.05(0.05) & 15.36(0.03) & 15.01(0.02) & 14.82(0.03) & LT+RATCam\\
2004 Feb 05 & 3040.92 &  30 & 16.27(0.04) & 15.52(0.02) & 15.05(0.01) & 14.79(0.03) & P60 \\
2004 Feb 06 & 3041.92 &  31 & 16.33(0.09) & 15.44(0.03) & 15.01(0.02) & 14.71(0.03) & P60 \\
2004 Feb 07 & 3042.92 &  32 & 16.23(0.06) & 15.49(0.04) & 15.02(0.02) & 14.77(0.03) & P60 \\
2004 Feb 08 & 3043.91 &  33 & 16.34(0.06) & 15.45(0.04) & 15.03(0.02) & 14.77(0.05) & P60 \\
2004 Feb 13 & 3048.90 &  38 & 15.92(0.03) & 15.17(0.04) & 14.89(0.02) & 14.69(0.03) & P60 \\
2004 Feb 15 & 3050.90 &  40 &         $-$ & 15.35(0.02) & 14.94(0.01) & 14.71(0.04) & P60 \\
2004 Feb 17 & 3052.89 &  42 & 16.27(0.02) & 15.32(0.02) & 14.98(0.02) & 14.68(0.04) & P60 \\
2004 Feb 18 & 3053.89 &  43 & 16.36(0.04) & 15.48(0.05) & 15.09(0.04) & 14.85(0.08) & P60 \\
2004 Mar 14 & 3078.84 &  68 & 16.43(0.01) & 15.31(0.03) & 14.91(0.01) & 14.61(0.02) & P60 \\
2004 Mar 23 & 3088.02 &  77 & 16.71(0.04) & 15.60(0.07) & 14.95(0.05) & 14.55(0.12) & P60 \\
2004 Apr 21 & 3116.97 & 106 & 16.97(0.01) & 15.76(0.02) & 15.27(0.01) & 14.89(0.02) & P60 \\
2004 Apr 24 & 3119.83 & 109 & 17.25(0.01) & 15.86(0.02) & 15.43(0.01) & 15.04(0.01) & P60 \\
2004 Apr 26 & 3121.83 & 111 & 17.15(0.02) & 16.13(0.04) &         $-$ & 14.98(0.03) & P60 \\
2004 Apr 27 & 3122.80 & 112 & 17.49(0.01) & 15.78(0.03) & 15.56(0.01) & 15.19(0.01) & P60 \\
2004 Apr 28 & 3123.79 & 113 & 17.41(0.01) & 16.27(0.02) & 15.66(0.01) & 15.23(0.01) & P60 \\
2004 Apr 29 & 3124.79 & 114 & 17.30(0.02) & 16.21(0.02) & 15.54(0.02) & 15.16(0.03) & P60 \\
2004 May 05 & 3130.84 & 120 & 18.11(0.04) &         $-$ &         $-$ &         $-$ & P60 \\
2004 Jun 03$^a$ & 3160.49 & 150 & 19.04(0.04) & 17.68(0.05) & 16.93(0.03) & 16.44(0.01) & WHT+AUX \\
2004 Aug 30 & 3248.47 & 238 & 19.47(0.12) & 18.25(0.04) & 17.62(0.04) & 17.29(0.03) & LT+RATCam \\
2004 Sep 23 & 3271.65 & 261 & 19.95(0.03) & 19.02(0.04) &         $-$ & 17.60(0.04) & {\it HST}+ACS \\
        \hline
      \end{tabular}\\
    \end{center}
    {\scriptsize NOTE: Figures in brackets give the statistical errors associated with the magnitudes.\\
      $^a$Observers were R. \O stensen and C. R. Benn. LT = 2.0-m Liverpool Telescope, La Palma. P60 = Robotic Palomar 60-inch telescope, Palomar. WHT = 4.2-m William Herschel Telescope, La Palma.}
  \end{minipage}
  \label{tab:phot}
\end{table*}

\begin{figure}
  \begin{center}
    \epsfig{file = 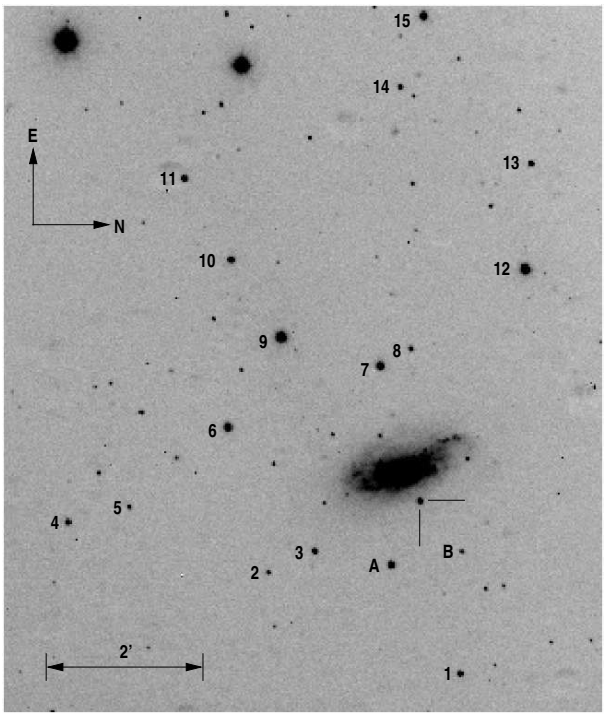,width = 80mm}
    \caption{Finder chart for SN~2004A. The letters denote the stars that were calibrated from the WHT-AUX images. The numbers denote stars that were calibrated using stars~A and~B.}\label{fig:finder}
  \end{center}
\end{figure}

The Johnson-Cousins {\it BVRI} magnitudes, in all epochs, were obtained using the aperture photometry task within the {\sc iraf} package {\sc daophot}, and were calibrated using differential photometry. Unfortunately there were no standard fields taken on the same nights as the SN data, except with those taken on 2004 June 3, with the AUX. The AUX, with its 1024$\times$1024 TEK CCD, has an unvignetted, circular field diameter of 1.8 arcmin, which meant that only stars~A and~B, from the SN~2004A field in Fig.~\ref{fig:finder}, fell within the aperture. Because of the highly variable quality of the images, {\em both} stars~A and~B were not well observed in every epoch (mostly because of B being significantly fainter than the SN). We could have used only these two stars but decided instead to employ a two step process. We first calibrated stars~A and~B from the AUX images, and then used them to calibrate the other stars in the field, using the night with the best quality images, to create a catalogue from which the SN could then be calibrated. To this end stars~A and~B, and SN 2004A, were calibrated using the standard stars, SA~107 626 and 627. The night with the best quality images, 2004 April 24, was then chosen to represent our `standard field'. The remainder of the stars, that were present and usable in all epochs, were calibrated using stars~A and~B. The stars calibrated in this way are numbered in Fig.~\ref{fig:finder} and their {\it BVRI} magnitudes are given in Table \ref{tab:cat}, where the numbers in brackets are the statistical errors. The magnitudes of the `standard' stars, A--B and 1--15, are the simple average of the individual values determined from each calibration star from two different images, and the error is the standard deviation.

\begin{table*}
  \caption[]{Photometry of the stars in the field surrounding SN~2004A. The stars are labelled in the same way as in Fig.~\ref{fig:finder}. Stars~A and~B were calibrated from the WHT-AUX images and 
the other numbered stars were then calibrated using stars~A and~B. The figures in brackets give the statistical errors associated with the magnitudes.}
    \begin{center}
      \begin{tabular}{rrrrr} \hline
        ID & $B$ & $V$ & $R$ & $I$ \\
        \hline
        A & 15.959(0.008) & 15.195(0.018) & 14.736(0.035) & 14.322(0.017) \\
        B & 18.081(0.019) & 17.560(0.015) & 17.267(0.034) & 16.943(0.021) \\
        1 & 16.546(0.022) & 15.901(0.066) & 15.541(0.028) & 15.213(0.017) \\
        2 & 17.664(0.022) & 17.104(0.066) & 16.791(0.028) & 16.548(0.017) \\
        3 & 16.283(0.022) & 15.563(0.066) & 15.169(0.028) & 14.842(0.017) \\
        4 & 16.527(0.022) & 15.898(0.066) & 15.565(0.028) & 15.275(0.017) \\
        5 & 17.808(0.022) & 16.989(0.066) & 16.535(0.028) & 16.173(0.017) \\
        6 & 14.914(0.022) & 14.242(0.066) & 13.889(0.028) & 13.562(0.017) \\
        7 & 14.558(0.022) & 13.966(0.066) & 13.639(0.028) & 13.355(0.017) \\
        8 & 16.971(0.022) & 16.305(0.066) & 15.950(0.028) & 15.696(0.017) \\
        9 & 13.868(0.022) & 13.016(0.066) & 12.548(0.028) & 12.157(0.017) \\
        10 & 15.756(0.022) & 15.135(0.066) & 14.800(0.028) & 14.483(0.017) \\
        11 & 16.093(0.022) & 15.140(0.066) & 14.620(0.028) & 14.165(0.017) \\
        12 & 14.165(0.022) & 13.127(0.066) & 12.575(0.028) & 12.111(0.017) \\
        13 & 16.675(0.022) & 15.999(0.066) & 15.624(0.028) & 15.309(0.017) \\
        14 & 16.739(0.022) & 16.012(0.066) & 15.614(0.028) & 15.270(0.017) \\
        15 & 14.583(0.022) & 14.050(0.066) & 13.757(0.028) & 13.508(0.017) \\
        \hline
      \end{tabular}
    \end{center}
  \label{tab:cat}
\end{table*}

\begin{figure}
  \begin{center}
    \epsfig{file = 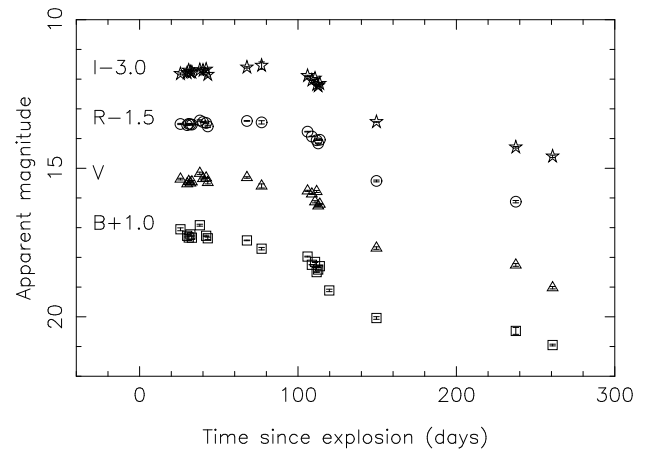, angle = -90, width = 82mm}
    \caption{{\it BVRI} light curves of SN~2004A, which have been arbitrarily shifted in magnitude for clarity. There is 
a clear and well developed plateau phase in each filter lasting around 80-100 days from the estimated explosion epoch. 
}\label{fig:BVRI}
  \end{center}
\end{figure}

The catalogue in Table \ref{tab:cat} was used to calibrate all the epochs in Table \ref{tab:phot}, except those from the {\it HST} and, as we have discussed, the AUX. The {\it BVRI} light curves are plotted in Fig.~\ref{fig:BVRI}. The `standard' stars were visible in the majority of the epochs although the field size and pointing of the LT meant that only stars A, B, 3, 6, 7, 8, 9 and 10 were used for the night of 2004 February 1 and stars  A, B, 1, 7 and 8 for the night of 2004 August 30. The SN magnitudes are the simple average of the individual values determined from each calibration star, where any outliers that were greater than $2\sigma$ away from the mean were omitted. The statistical error, shown in brackets, is either the standard deviation or the combined error from each of the estimates, whichever was the greater.

In the course of calibrating our `standard' field, the SN magnitudes for 2004 April 24, were also calibrated using only stars~A and~B. The difference between these magnitudes and those obtained using the catalogue, gave us an estimate of the systematic error introduced by the calibration, including the transformation from our `standard' system to the AUX system. We therefore estimate the calibration error, for bands {\it BVRI}, to be 0.01, 0.12, 0.03 and 0.07, respectively.

The observations are from four different instrumental set-ups and filter systems; the LT, the P60 before 2004 March 14, the P60 after 2004 March 14, and the WHT. There are two instrumental set-ups for the P60 as its CCD was changed on 2004 March 14. The colour transformation plots of the instrumental magnitudes, from both the LT and the P60 pre-2004 March 14 systems, compared to the intrinsic colours of the `standard' stars suggested that it was unnecessary to apply a colour correction to any of the {\it BVRI} magnitudes, to transform the instrumental magnitudes to our `standard' system. We would, however, add caution to the {\it RI} magnitudes from the LT, as these are Sloan $r\arcmin i\arcmin$ magnitudes. The transformation from the $r\arcmin i\arcmin$ to the {\it RI} system was found to be problematic by \citet{2005MNRAS.359..906H}. When the transformations were applied to the field stars the {\it RI} results were reliable, however when the same transformations were applied to the SN they gave very unsatisfactory results. The authors suggested that the strong emission line SN spectrum produced systematic differences in the colour terms compared to the stellar SEDs. We believe that it is not advantageous to apply a colour correction to the Sloan $r\arcmin i\arcmin$ magnitudes here and we have retained them as pseudo-{\it RI} magnitudes. The two LT {\it RI} points are not deviant from the light curves in Figs.~\ref{fig:BVRI} and \ref{fig:04A99em}, and the point in the tail also fits very well to both the other SN~2004A data and the extrapolated SN~1999em light curves, lending credence to this assumption. The RMS of the residuals for each system, assuming no colour correction, are shown in Table \ref{tab:rms}. It is difficult to determine if a colour correction is needed to transform our `standard' system to the Landolt system, because both the AUX images and our `standard' field were only calibrated using two standards. The transformation plots, however, did not suggest that one was required.

\begin{table}
  \caption[]{RMS residuals assuming no colour transformations between the instrumental magnitudes and the intrinsic colours.}
    \begin{center}
      \begin{tabular}{lrrrr} \hline
        Transformation & $B$ & $V$ & $R$ & $I$ \\
        \hline
        AUX$\rightarrow$Landolt & 0.01 & 0.01 & 0.01 & 0.01 \\
        `standard'$\rightarrow$AUX & 0.02 & 0.05 & 0.01 & 0.02 \\
        LT$\rightarrow$`standard'& 0.05 & 0.04 & 0.03 & 0.03 \\
        P60 (pre-040314)$\rightarrow$`standard' & 0.04 & 0.04 & 0.02 & 0.03 \\
        \hline
      \end{tabular}
    \end{center}
    \label{tab:rms}
\end{table}

The error introduced to the LT and the P60, pre-2004 March 14,
magnitudes by assuming there was no colour correction, was estimated
from the best linear fits to the transformation plots. 
These errors were comparable to the values for the gradient and
offset, and often larger, and the reduced-$\chi^2$ values
were comparable to those for the null hypothesis, hence there is 
no value in applying them. As there are only
two standard stars each for the transformations from our `standard'
system to the Landolt system, we were unable to estimate an error in
this way. However, we have already estimated an error for the
transformation from our `standard' system to the AUX system, leaving
only the AUX to Landolt transformation, where we have used the RMS
residual value to represent the error. We have added these errors in
quadrature to the statistical error, hence this uncertainty is
accounted for throughout the analysis in the forthcoming sections. We
show that the major conclusions of this paper are not critically
dependent on these uncertainties.

The error introduced by the difference between the Cousins {\it RI} and the Sloan $r\arcmin i\arcmin$ filters was quantified using spectrophotometry, using the {\sc iraf} package {\sc synphot} within {\sc stsdas}, applied to the only spectrum of SN~2004A. Using the rescaled spectrum (see Section~\ref{sec:spec}) the differences $R-r\arcmin$ and $I-i\arcmin$ were $-0.06$ and 0.01, which are comparable to the estimated errors for $R$ and well within the errors in $I$. SN~1999em at a similar phase also differed by consistent amounts. However, the differences between the {\sc synphot} and LT Sloan $r\arcmin i\arcmin$ filters were not addressed. We do not have spectra from phases close to the LT photometry epochs so we used the spectra of SN~1999em to investigate these as they gave consistent results at around 40\,d. At a phase of 23\,d the differences were $-0.02$ and 0.03, and at 165\,d they were $-0.01$ and $-0.05$, respectively. The errors estimated for the LT are large enough to account for any error introduced by the difference in the Cousin and Sloan filters. In any case there are only two LT epochs and the conclusions of this paper are not reliant on these points.


Since the data in this paper was reduced, calibrated and analysed, the Sloan Digital Sky Survey (SDSS) data release~4 (DR4) was made public. As a check the photometry was redone using the standard sequence given in Table~\ref{tab:sdss}. Star~A, which was one of the stars used to calibrate the `standard' field, differs by: 0.12, 0.07, 0.06 and 0.04 mag in {\it BVRI}, respectively. The SN photometry calibrated using the SDSS catalogue (Table~\ref{tab:sdss}) is fainter over all the epochs and bands, with differences, averaged over all the epochs, of: 0.14, 0.12, 0.08 and 0.02 for each band. The differences in the light curves are however minimal. The calibration errors for using the catalogue in Table~\ref{tab:cat}, as opposed to just stars~A and~B were: 0.01, 0.12, 0.03 and 0.07, for bands {\it BVRI} respectively. These errors were added in quadrature to the errors derived from the LT and the P60, pre-2004 March 14, colour transformation equations, so will adequately cover the differences in the photometry. The main results on which the SDSS catalogue will have an effect are the standard candle method (SCM) distance estimate, and consequently the nickel mass. However, the SCM distance was found not to change significantly, increasing the overall distance to NGC~6207 by only 0.2\,Mpc. Therefore, the analysis and conclusions of this paper are the same irrespective of the catalogue that is used. There are errors associated with the transformation of the SDSS stellar magnitudes to the Johnson--Cousins filter system, so the SDSS catalogue in Table~\ref{tab:sdss} is probably not any more reliable than the catalogue used here. Hence the photometry of SN~2004A from the SDSS catalogue, because of the errors in the transformation of the catalogue, is not any more reliable than the photometry presented here. It's agreement is reassuring though and makes the further analysis all the more robust.

\begin{table}
  \caption[]{Johnson-Cousins BVRI magnitudes, from SDSS DR4, for selected stars in the field of SN~2004A, using the same sequence as Figure \ref{fig:finder}.}
    \begin{center}
      \begin{tabular}{rrrrr} \hline
        ID & $B$ & $V$ & $R$ & $I$ \\
        \hline
 A & 16.083 & 15.264 & 14.796  & 14.364 \\
 3 & 16.422 & 15.684 & 15.263 & 14.885 \\
 6 & 15.039 & 14.349 & 13.955 & 13.575 \\
 7 & 14.748 & 14.136 & 13.787 & 13.381 \\
 9 & 16.338 & 13.941 & 12.560 & 12.090 \\
10 & 15.891 & 15.257 & 14.897 & 14.510 \\
12 & 16.437 & 14.002  & 12.599 & 12.046 \\
15 & 14.705 & 14.175 & 13.874 & 13.545 \\
        \hline
      \end{tabular}
    \end{center}
  \label{tab:sdss}
\end{table}

\subsection{Spectroscopy of SN~2004A}\label{sec:spec}

\begin{figure}
  \epsfig{file = 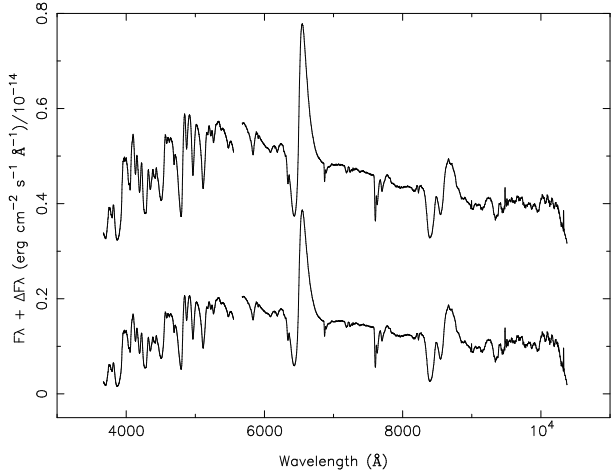, width = 80mm}
  \caption{The original flux calibrated spectrum (bottom), showing a slightly flatter 
continuum below 6\,000\AA, and the same spectrum with the linear scale function applied (top).}\label{fig:spec}  
\end{figure}

An optical spectrum of SN~2004A was obtained on 2004 Feb 18.65 {\sc ut}, at
a phase of $\sim$43\,d (see Section~\ref{sec:epoch}), using the LRIS
instrument on Keck. The details of the spectrum are given in Table~\ref{tab:spec}. The spectrum was reduced using standard routines
within {\sc iraf}. The frames were debiased, flat-fielded and
extracted and then wavelength calibrated using Cu-Ar and Cu-Ne lamp
spectra. The wavelength calibration was checked by determining the
positions of the night sky lines and small adjustments were made. The
spectra were then flux calibrated using spectrophotometric flux
standards observed with the same instrumental setup. The slit-width
employed was 0.7 arcsec, hence the flux calibration is unlikely to
give an accurate absolute scale. The continuum of the resulting
spectra was visibly flat in the region lower than 6\,000\AA\
(see Fig.~\ref{fig:spec}, bottom),
possibly indicating a problem with the flux calibration. We therefore
used the {\it BVRI} photometry to adjust the flux calibration. 
Spectral {\it BVRI} magnitudes were calculated from
the spectra using the {\sc iraf} package {\sc synphot} within {\sc
stsdas}. These magnitudes were compared to the observed photometric
magnitudes. In order to match the photometric magnitudes a linear 
scaling function (linear in flux and wavelength) was calculated and
applied to the spectrum. 
The original flux calibrated spectrum and the scaled spectrum
are shown in Fig.~\ref{fig:spec}. The spectrum was read into the
spectral analysis program {\sc dipso} \citep{SUN50.24} for further
analysis.

\begin{figure}
  \epsfig{file = 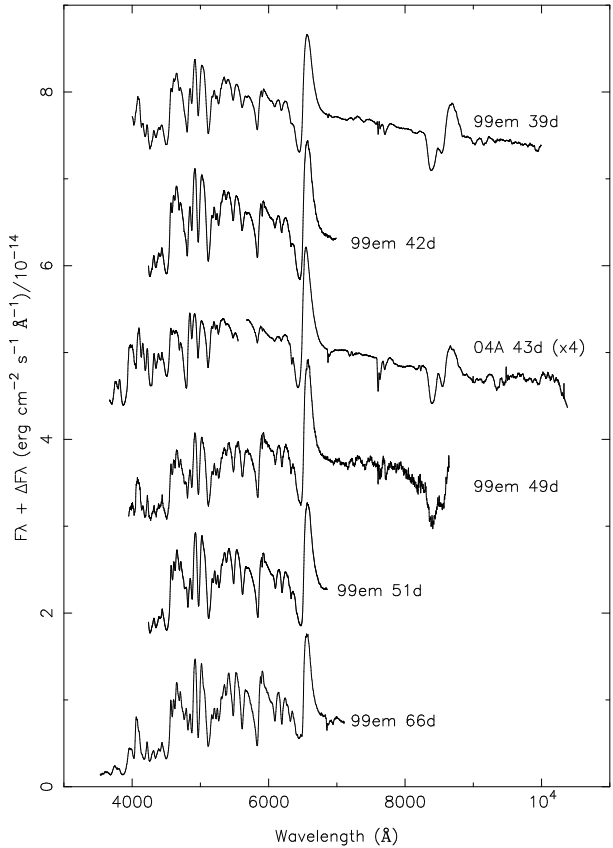, width = 80mm}
  \caption{The scaled spectrum of SN~2004A compared with the spectra of SN~1999em 
\citep{2002PASP..114...35L,2001ApJ...558..615H}
over the range of plausible phases for SN~2004A, from the explosion epoch. The spectrum of SN~2004A has been further scaled for clarity.}\label{fig:s04A99em} 
\end{figure}

The spectrum of SN~2004A is compared to that of the well observed
SN~1999em, in Fig.~\ref{fig:s04A99em}, over the plausible phases
inferred from the observations of Itagaki
\citep{2004IAUC.8265....1N}. The spectrum looks similar to those of
SN~1999em at a similar phase and shows that the $\chi^2$-fitting
method (Section~\ref{sec:epoch}) may have over-estimated the explosion
epoch by a few days. The spectrum shows P-Cygni profiles and broad
spectral features, indicative of the high velocities of the SN. The
spectrum will be available through the {\sc
suspect}\footnote{http://bruford.nhn.ou.edu/$\sim$suspect/} website.

\begin{table}
\caption[]{Details of optical spectrum of SN~2004A}\label{tab:spec}
  \centering
  \begin{tabular}{ll}
    Date & 2004 Feb 18\\
    JD (+245\,0000) & 3054.15\\
    Phase & 43\,d\\
    Range &3373--10380 \AA\\
    Resolution & 10.6 \AA\\
    Telescope/Instrument & Keck/LRIS\\
    Observer & Smartt/Maund
  \end{tabular}
\end{table}

\subsection{{\it HST} Observations of NGC~6207}\label{sec:hst}

Observations of NGC~6207, before the explosion of SN~2004A, were taken with WFPC2 on board {\it HST} on two separate occasions through the filters F606W, F814W and F300W, as part of programs SNAP9042 and GO8632 (all details are given in Table~\ref{tab:tab1}). Further {\it HST} imaging was acquired with the Wide Field Camera (WFC) of the Advanced Camera for Surveys (ACS). SN~2004A was observed at an epoch $\sim$261\,d after explosion using three filters F435W, F555W and F814W as part of program GO9733. These observations were acquired in order to determine a value of the reddening towards stars near the SN and to determine the exact location of the SN with respect to nearby stars. These images allowed the precise location of the SN to be determined on the pre-explosion observations. The target was placed close to the centre of chip WF1. 

The on-the-fly re-calibrated (OTFR) ACS images were obtained from the Space Telescope European Coordinating Facility archive. Photometry was conducted on these frames using the new custom built ACS PSF-fitting photometry modules in the package {\sc dolphot}\footnote{http://purcell.as.arizona.edu/dolphot/} \citep{2000PASP..112.1383D}. This incorporates PSF fitting with model ACS PSFs, charge transfer efficiency (CTE), aperture corrections and transformation between the flight system magnitudes and standard Johnson-Cousins {\it BVI} filter bands. Stars, suitable for the brightest supergiants method (BSM) distance determination technique (see Section~\ref{sec:BSGs}), were selected from the photometry output using the {\sc dolphot} object type classification scheme. Objects that were classified by {\sc dolphot} as extended, blended or containing bad pixels were discarded, as they were probably blended stars or stars in crowded areas with PSF fit $\chi^2>2.5$.

The WFPC2 pre-explosion images were retrieved from the Space Telescope Science Institute archive, and calibrated via the OTFR pipeline. These observations were made in three bands F300W, F606W and F814W. The site of SN 2004A was located on the WF3 chip in all WFPC2 observations, which has a resolution of 0.1 arcsec per pixel. Aperture photometry was conducted using the {\sc iraf} package {\sc daophot} and PSF photometry performed using the {\sc hstphot} package \citep{2000PASP..112.1383D}. {\sc hstphot} includes corrections for chip-to-chip variations, CTE and transformations from WFPC2 instrumental magnitudes to the standard Johnson-Cousins magnitude system.

\begin{table*}
  \caption[]{Summary of {\it HST} observations}
  \begin{center}
    \begin{tabular}{llllrll} \hline
      & Date &  Filter & Dataset & Exposure (s) & Instrument & Program ID\\
      \hline
      Pre-explosion  & 2000 Aug 03  &   F300W  &  U67GA101B  &   1000  &   WFPC2  &  GO8632\\
      & 2001 Jul 02  &   F606W  &  U6EAD003B  &    460  &   WFPC2  &  SNAP9042\\
      & 2001 Jul 02  &   F814W  &  U6EAD001B  &    460  &   WFPC2  &  SNAP9042\\
      \\
      Post-explosion & 2004 Sept 23   &  F435W &   J8NV02010  &    1400  &   ACS/WFC  & GO9733\\
      & 2004 Sept 23   &  F555W &   J8NV02020  &    1510  &   ACS/WFC  & GO9733\\
      & 2004 Sept 23   &  F814W &   J8NV02030  &    1360  &   ACS/WFC  & GO9733\\
      \hline
    \end{tabular}
  \end{center}
  \label{tab:tab1}
\end{table*}

\section{Analysis of the evolution of SN~2004A}

\subsection{Explosion epoch}\label{sec:epoch}

\begin{figure*}
\begin{center}
\epsfig{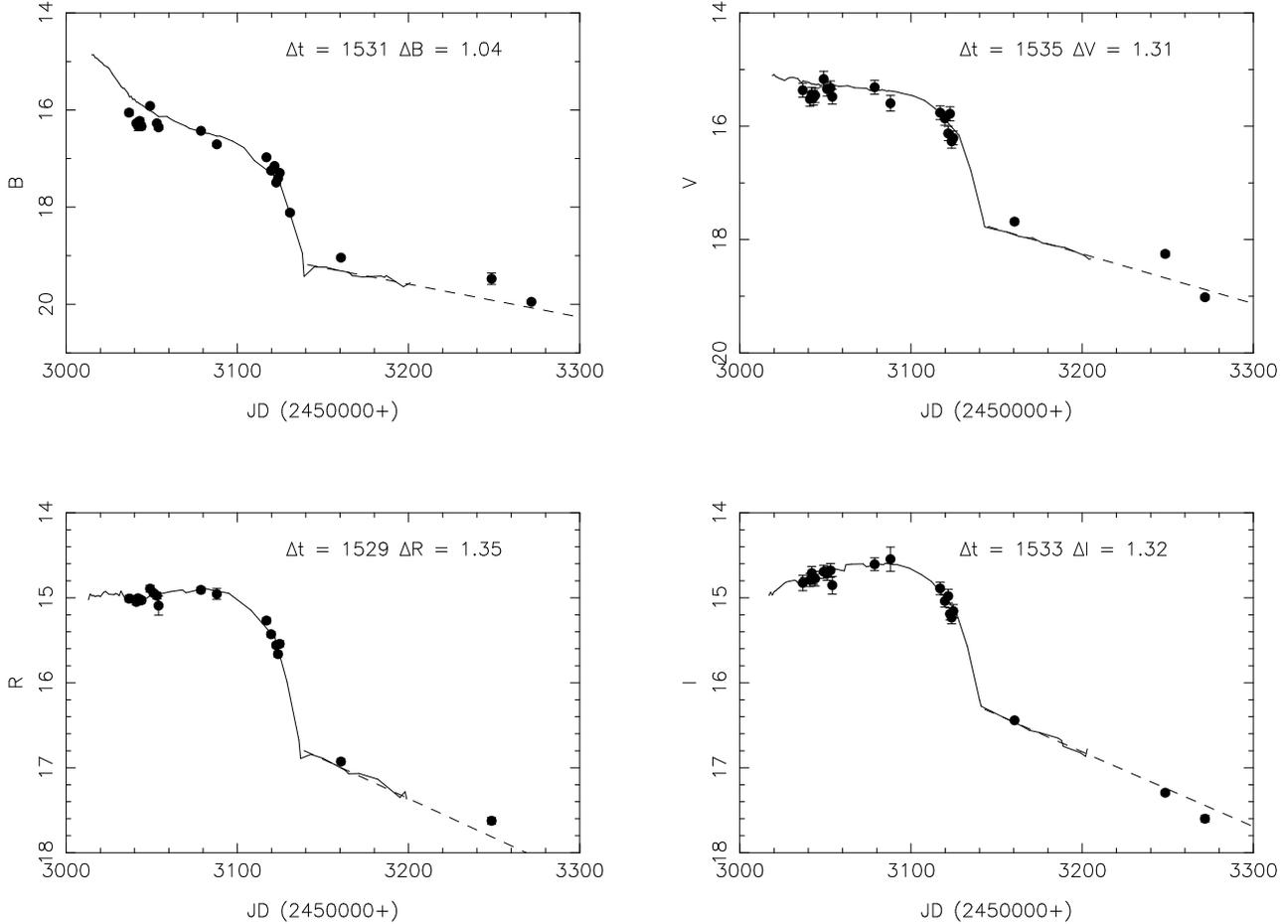}
\caption{{\it BVRI} light curves of SN~2004A, shown with the filled
circles, over-plotted with the best fit light curve of SN~1999em, from
\citet{2001ApJ...558..615H}, shown with the solid line (which has
been shifted by $\Delta t$ days and $\Delta m$ magnitudes). The
reduced-$\chi^2$ of the fit for the comparison of the {\it BVRI} light
curves are: 23.10, 2.91, 2.30 and 0.74. The dashed line represents the
best linear fit to the data in the tail of SN~1999em.}
\label{fig:04A99em}
\end{center}
\end{figure*}

Using a similar method to \citet{2005MNRAS.359..906H}, the light curve of SN~2004A was compared to that of SN~1999em \citep{2001ApJ...558..615H}, which is the most fully studied normal SN II-P to date, to obtain an estimate of the explosion epoch. This was accomplished using a $\chi^2$-fitting algorithm to adjust the time and apparent magnitude of the `model' light curve of SN~1999em, to find the best fit to the data points of SN~2004A. The SN~2004A errors used in this analysis were the statistical plus the systematic error discussed in Section \ref{sec:phot}. The last two points in the tail of SN~2004A were not included in the fit as there was insufficient data from the SN~1999em light curve, as can be seen from Fig.~\ref{fig:04A99em}. The {\it BVRI} best fits are shown in Fig.~\ref{fig:04A99em} with the shift in time, $\Delta t$, and apparent magnitude, $\Delta m$, inset in each figure. The results from the $\chi^2$-fitting algorithm are shown in Table~\ref{tab:chi2LC}.

\begin{table}
\begin{center}
\caption{Results from the $\chi^2$-fitting algorithm which adjusts the time, $\Delta t$, and apparent magnitude, $\Delta m$, of the 'model' light curve (SN~1999em) to find the best fit to the data points of SN~2004A, where $\nu$ $=$ number of data points $-$ number of degrees of freedom.}
\label{tab:chi2LC}
\begin{tabular}{lrrrr}\hline
Filter & Reduced-$\chi^2$ & $\nu$ & $\Delta t$ (days) & $\Delta m$\\
\hline
$B$ & 23.10 & 18 & 1531(3) & 1.04(0.08)\\
$V$ &  2.91 & 18 & 1535(3) & 1.31(0.08)\\
$R$ &  2.30 & 17 & 1529(3) & 1.35(0.04)\\
$I$ &  0.74 & 18 & 1533(3) & 1.32(0.04)\\[4pt]
{\it BVRI} &    &      & 1532(3) & \\
\hline
\end{tabular}
\end{center}
\end{table}

The reduced-$\chi^2$ values are exceedingly large for the $B$ band and
is indicative of a poor fit. The fit by eye looks quite poor as there
is a large scatter about the light curve. The statistical and
systematic errors are small in the $B$ band, but the scatter suggests
that this may have been underestimated. The light curves, although
they have the same general shape, are different in the early part of
the $B$-band light curve and in the tail. The early SN~2004A light
curve appears to be flatter and the tail slightly more luminous,
increasing the $\chi^2$ value. The $V$-band fit also has a reasonably
high reduced-$\chi^2$, although it is lower than that of the $B$-band
fit. There is a reasonably large scatter around the light curve,
however the shape of the light curves are more consistent than in the
$B$-band. The {\it RI} bands visually appear to be good fits and their
reduced-$\chi^2$ values are correspondingly lower, although the value
for the $R$ band is still on the high side. The points in the SN~2004A
tail are also fitted well by the SN~1999em light curve and the later
points, omitted from the fit, are also in agreement with the
extrapolated light curve, shown with the dashed line. There is some
scatter in the points compared to SN~1999em, which is due to the 
variable quality of the imaging for this event and very unlikely to be
due any intrinsic brightness variations. 

The reduced-$\chi^2$ values were too large in the {\it BVR} bands to be
able to do a sensible error analysis using the confidence limits. We
estimated the errors instead, in all bands, by first rounding $\Delta
t$ to the nearest day and then fitting the light curves by eye to
obtain a realistic range of values for $\Delta t$ and $\Delta
m$. These ranges are the errors quoted, in brackets, in Table~\ref{tab:chi2LC}. The $I$-band fit was sufficiently good to estimate
an error from the confidence limits. Using the 2$\sigma$ confidence
limit, we estimate, $\Delta t(I) = 1532.5^{+2.5}_{-3.0}$, rounded to
the nearest half day, and $\Delta I = 1.32^{+0.04}_{-0.08}$, which is
in agreement with what was found by eye.

The explosion epoch of SN~2004A, using this method, was estimated to
be JD $245\,3011\pm3$, which corresponds to 2004 January 6. This was
achieved using the simple average of $\Delta t$, for {\it BVRI}, and the
explosion epoch of SN~1999em, which was estimated to be JD $245\,1478.8\pm0.5$ by \citet{2001ApJ...558..615H}, within two days of
\citet{2003MNRAS.338..939E}. The error in the explosion epoch was
estimated from the errors in the weighted average of $\Delta t$ and in
the SN~1999em explosion date. The error of $\pm 3$\,d is the error on the fit and does not reflect the systematic error that is introduced by assuming that both SN~2004A and SN~1999em are intrinsically the same. We can, however, estimate a more appropriate error from observations. We can put a hard limits on the earliest and latest possible explosion date of
SN~2004A by using the observations of K. Itagaki
\citep{2004IAUC.8265....1N}. The SN was discovered by Itagaki
\citep{2004IAUC.8265....1N} on Jan. 9.84 {\sc ut} and it was not seen
on Itagaki's previous observations of 2003 December~27, which have a
limiting magnitude of 18. The explosion epoch found by the
$\chi^2$-fitting method is consistent with this limit, which 
gives us a robust uncertainty range for the explosion 
epoch of  JD $245\,3011^{+3}_{-10}$.

\subsection{Reddening estimate towards SN~2004A}\label{sec:red}

\subsubsection{Reddening towards the neighbouring stars}

{\it BVI} ACS photometry (see Section~\ref{sec:hst}) was used to estimate the reddening towards SN~2004A. \bv\ and \vi\ colours of stars within 6 arcsec of SN~2004A were compared with the intrinsic supergiant colour sequence of \citet{2000asqu.book.....DL}. The reddening was calculated using a $\chi^2$-minimisation of the displacement of the stars from the intrinsic supergiant colour sequence, for a range of values of \ebv.  The reddening vector, in the \bv/\vi\ colour plane, assumed the reddening laws of \citet{1989ApJ...345..245C} with $R_V = 3.1$. Using this method the reddening was estimated as $\ebv=0.06\pm0.03$, which corresponds to $A_U = 0.29 \pm 0.15$, $A_V = 0.19 \pm 0.09$, $A_I = 0.09 \pm 0.05$.

\subsubsection{Reddening and the colour evolution of SN~2004A}

The reddening towards SNe can also be estimated from a comparison
between the colour evolutions of the SN in question and another well
studied SN, for which the reddening is known to some degree of
accuracy. This method assumes that SNe~II-P all reach the same
intrinsic colour towards the end of the plateau phase. This is based
on the assumption that the opacity of SNe~II-P is dominated by
electron scattering, therefore the SN should reach the temperature
of hydrogen recombination at the end of the plateau phase
\citep{1996ApJ...466..911E,2004mmu..sympE...2H}. 
Unfortunately the photometry was not accurate enough to estimate
an independent reddening towards the
SN, although we could ascertain whether the estimate from the
surrounding stars and the colour evolution were consistent.

\begin{figure}
  \centering
  \epsfig{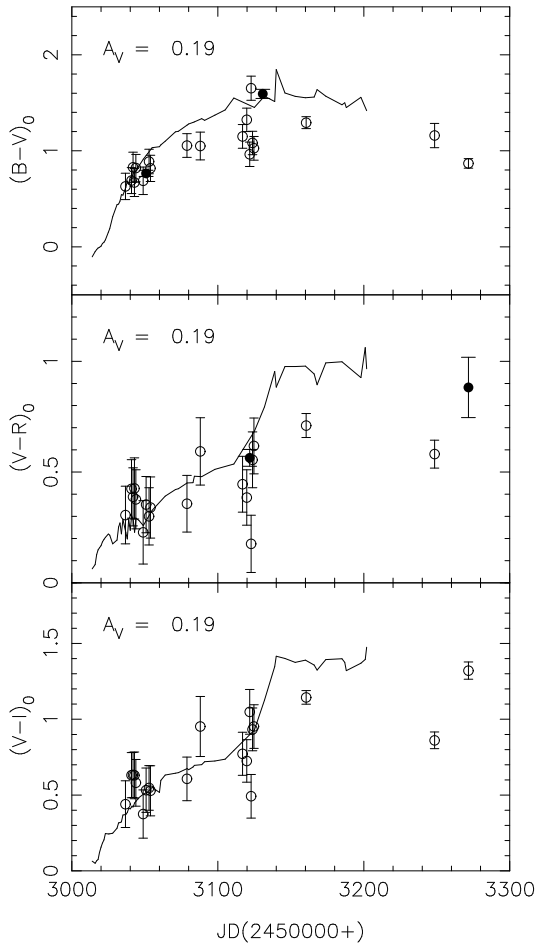}
  \caption{Colour evolution of SN~2004A (points) compared with that of SN~1999em (solid line)
for $A_V = 0.19$.}
  \label{fig:c04A99em}
\end{figure}

We compared the colour evolution of SN~2004A with that of SN~1999em, following a similar approach to \citet{2005MNRAS.359..906H}, who adopted a reddening of $\ebv = 0.075\pm0.025$ for SN~1999em \citep{2000ApJ...545..444B}. The colour curves of SN~1999em were firstly dereddened using this value and were then shifted in time using the weighted average of $\Delta t$ discussed in Section~\ref{sec:epoch}. The colour curves of SN~2004A were then dereddened using the estimate from the neighbouring stars estimated here. The comparison of the colour evolutions are shown in Fig.~\ref{fig:c04A99em} and the reduced-$\chi^2$ values are listed in Table~\ref{tab:cchi2}. The $\chi^2$-fit was restricted to JD $<245\,3120$, the end of the plateau.
\begin{table}
\caption{Reduced-$\chi^2$ values from the comparison of the colour evolutions of SNe~2004A and 1999em, for an extinction of $\ebv=0.06\pm0.03$ for SN~2004A, estimated from the neighbouring stars.}\label{tab:cchi2}
  \centering
  \begin{tabular}{ccc}
    \hline
    & reduced-$\chi^2$ & $\nu$\\
    \hline
    \bv   &  2.035 & 12\\
    \vr   &  0.940 & 13\\
    \vi   &  0.969 & 13\\
    \hline
  \end{tabular}\\
\end{table}

The reduced-$\chi^2$ values are indicative of a good fit for the \vr\
and \vi\ colours, although the error bars are quite large. The quoted
uncertainties
are a combination of the statistical errors from the photometry and
the errors associated with the calibration discussed in
Section~\ref{sec:phot}. The fits show that, even though we could not
obtain an independent reddening estimate, the colour evolution is
consistent with an extinction of $\ebv = 0.06\pm0.03$. As in
\citet{2004mmu..sympE...2H} and \citet{2005MNRAS.359..906H} we find
that the \bv\ is not as well behaved and has a higher $\chi^2$
value. Although the \bv\ fit is poorer, it is still consistent with
the reddening towards the neighbouring stars.

\subsection{Expansion velocity}\label{sec:expvel}

\begin{figure}
\centering
\epsfig{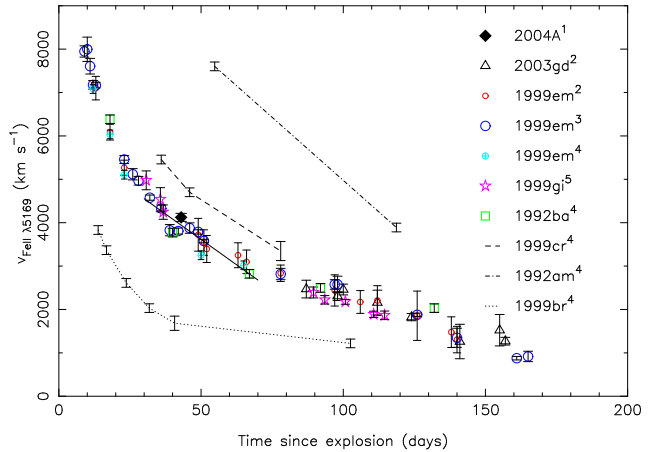}
\caption{Velocity evolution of SN~2004A compared with other similar SNe~II-P (1992ba, 1999em, 1999gi and 2003gd: plotted points) and contrasting SNe~II-P (1999cr, 1992am and 1999br: dashed/dotted curves). The solid line is the best fit to the similar SNe~II-P velocities between 30 and 70\,d. The superscripts in the figure denote the source of the velocity measurements: (1) this paper, (2) \citet{2005MNRAS.359..906H} (3) \citet{2002PASP..114...35L}, (4) \citet{2001ApJ...558..615H} and (5) \citet{2002AJ....124.2490L}.}
\label{fig:vSNeIIP}
\end{figure}

We have used the method of \citet{2005MNRAS.359..906H}, which was found to be consistent with other methods \citep{2001ApJ...558..615H,2002PASP..114...35L,2003MNRAS.338..939E}, to measure the expansion velocity of the ejecta of SN~2004A from our spectrum at 43\,d post-explosion. The expansion velocity was measured from the minimum of the blue-shifted absorption trough of the Fe {\sc ii} $\lambda$5169 line. The absorption trough was fitted by three Gaussians using the Emission Line Fitting package ({\sc elf}) within the {\sc starlink} spectral analysis package {\sc dipso}. The minimum was found from the three Gaussian fit and the error was estimated from the difference between this and a single Gaussian fit. We found the expansion velocity, of the ejecta of SN~2004A at 43\,d, to be $v = 4123\pm93$\,\kms, using the NASA/IPAC Extragalactic Database\footnote{http://nedwww.ipac.caltech.edu/} (NED) recessional velocity of 852\,\kms\ for NGC~6207. A comparison of the expansion velocities of SN~2004A and other \mbox{SNe II-P} is shown in Fig.~\ref{fig:vSNeIIP}. The SN~2004A expansion velocity is consistent with other normal SNe II-P, as we would expect from the appearance of the spectral features in Fig.~\ref{fig:s04A99em}.

\section{Distance estimates to NGC~6207}

\subsection{Type II-P standard candle method distance estimate}\label{sec:SCM}

Although SNe~II-P show a wide range of luminosities at all epochs, the velocities of their ejecta and their bolometric luminosities, during the plateau phase, are highly correlated. This correlation allowed \citet{2002ApJ...566L..63H} to formulate a Standard Candle Method (SCM) for \mbox{SNe~II-P}, which can be solved for the Hubble Constant provided a suitable distance calibrator is known. \citeauthor{2004mmu..sympE...2H} in two further papers \citep{2004mmu..sympE...2H,astro-ph/0309122} confirmed the SCM, using a sample of 24 SNe~II-P, and calibrated the Hubble diagram with known Cepheid distances to four of these SNe. \citet{astro-ph/0309122} calculated the Hubble constant, using $V$ and $I$ data, to be $H_0(V) = 75\pm7$ and $H_0(I) = 65\pm12$\,\hub. In this paper we use the weighted average of these results, $H_0 = 72\pm6$\,\hub, which is comparable to $H_0 = 71\pm2$\,\hub\ derived as part of the {\it HST} Key Project using SNe~Ia \citep{2001ApJ...553...47F}.

We used equations~(5) and~(6) from \citet{astro-ph/0309122} conversely to estimate the distance to SN~2004A:
 equations~(\ref{equ:V}) and~(\ref{equ:I}):

\begin{center}
\begin{equation}
\label{equ:V}
D(V) = \frac{10^{\frac{1}{5}(V_{50}-A_V+6.249(\pm 1.35)\log(v_{50}/5000)+1.464(\pm 0.15))}}{H_0}
\end{equation}
\begin{equation}
\label{equ:I}
D(I) = \frac{10^{\frac{1}{5}(I_{50}-A_I+5.445(\pm 0.91)\log(v_{50}/5000)+1.923(\pm 0.11))}}{H_0}
\end{equation}
\end{center}
where $V_{50}$, $I_{50}$ and $v_{50}$ are the $V$ and $I$ magnitudes, and the expansion velocity, in \kms, at a phase of 50\,d. The {\it VI} magnitudes and velocity of SN~2004A is known for a phase of 43\,d, so in order to use the SCM we were required to interpolate or extrapolate these quantities to 7d later.

There is quite a large scatter in the $V$ and $I$-band light curves so we fitted a straight line to the data between 20--80\,d to allow us to estimate $V_{50} = 15.41\pm0.17$ and $I_{50} = 14.68\pm0.13$. The errors are from the error in the fit and amply account for the errors in the photometry and the explosion date. Unfortunately we only have one spectrum, so we were unable to extrapolate the SN~2004A velocity using data from the SN alone. In order to estimate the velocity at 50\,d we used the locus of similar SNe~II-P velocities, which are consistent with that of SN~2004A (Fig.~\ref{fig:vSNeIIP}), to obtain an average velocity evolution. None of the SNe studied here deviate from this locus, so it is reasonable to assume that SN~2004A will also follow a similar evolution. We fitted a straight line to the velocities between 30--70\,d and used this to project the velocity to 50\,d, estimating $v_{50} = 3795^{+210}_{-502}$\,\kms. The errors are a combination of the errors in the fit, the explosion date and the velocity at 43\,d. The straight line fit is shown in Fig.~\ref{fig:vSNeIIP} by a solid line.

Using these parameters, the reddening calculated in Section~\ref{sec:red} and equations~(\ref{equ:V}) and~(\ref{equ:I}), we find $D(V) = 21.40^{+3.69}_{-4.89}$ and $D(I) = 20.67^{+2.89}_{-3.96}$\,Mpc, where the error is statistical and comes from combining the uncertainties of each parameter in the SCM equations. A straight average of these results gives a distance of $D = 21.0^{+4.1}_{-4.5}$\,Mpc, where the error is estimated from the limits of $D(V)$ and $D(I)$.

\subsection{Brightest supergiant distance estimate using {\it HST} photometry}\label{sec:BSGs}

The brightest supergiants distance method (BSM) uses the correlation between the average luminosity of a galaxy's brightest supergiants and the host galaxy luminosity. This average luminosity should be independent of the host galaxy luminosity for there to be no distance degeneracy \citep{1994MNRAS.271..530R}. We have used the method of \citet{1996AJ....111.2280S} to determine the magnitude of the brightest supergiants and the calibrations of both \citet{1994MNRAS.271..530R} and \citet{1994A&A...286..718K} to determine the distance. \citet{1996AJ....111.2280S} divided the supergiants of M74 into red and blue using their \vr\ colours, assuming $\vr = 0.5$ as the boundary between the red and blue supergiants. Taking into account M74's foreground reddening, this colour corresponds approximately to an F8 supergiant \citep[table 15.7]{2000asqu.book.....DL}. The brightest supergiants were then found from their luminosity functions by estimating the number of foreground stars in each bin. The brightest bin with an excess of supergiants will also statistically contain the brightest supergiants. This effectively removes the contamination from foreground stars, which cause the distance to be underestimated. In this work we have used an excess of 2$\sigma$ to indicate a significant detection. 

\begin{figure}
  \centering
  \begin{minipage}{0.5\textwidth}
    \centering
    \epsfig{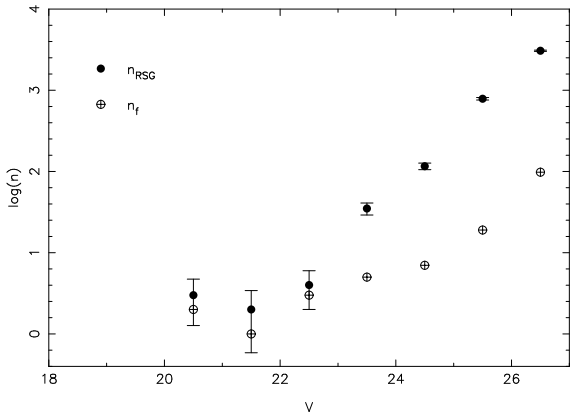}
  \end{minipage}\\[20pt]
  \begin{minipage}{0.5\textwidth}
    \centering
    \epsfig{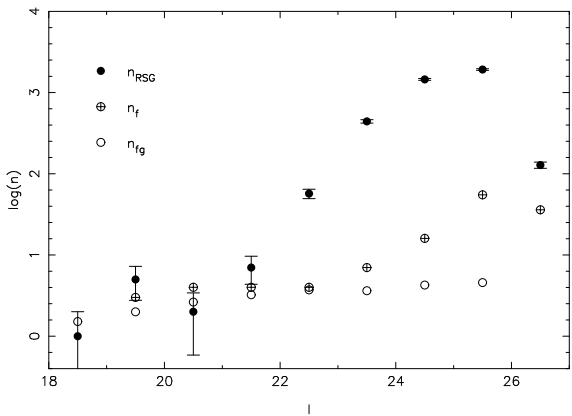}
  \end{minipage}
  \caption{{\it VI}-band luminosity functions for RSGs in NGC~6207, field stars adjacent to NGC~6207 and predicted stars from \citet{1981ApJS...47..357B} field~16. There are no predicted star counts in the $V$-band as they are not included in \citet{1981ApJS...47..357B}.}\label{fig:Rlfn}
\end{figure}

\begin{figure}
  \centering
  \epsfig{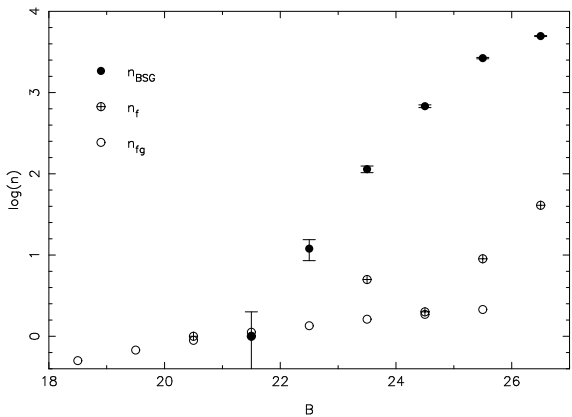}\caption{$B$-band luminosity function for BSGs in NGC~6207, field stars adjacent to NGC~6207 and predicted stars from \citet{1981ApJS...47..357B} field~16.}\label{fig:Blfn}
\end{figure}

The {\it HST} (ACS, see Section~\ref{sec:hst}) photometry was firstly
dereddened using the foreground extinction towards NGC~6207 from
\citet{1998ApJ...500..525S}, accessed through the NED
interface.
The supergiants were divided into red and blue using their \vi\
colours and the intrinsic colour of F8 type supergiants, which is
$\vi = 0.72$ from \citet[table 15.7]{2000asqu.book.....DL}. We have
assumed that supergiants with $\vi \ge 0.72$ are RSGs and supergiants with $\vi < 0.72$ are BSGs. The number of foreground stars were estimated using
\citet{1981ApJS...47..357B} field~16, which is the nearest field to
NGC~6207. As SN~2004A was placed close to the centre of chip WF1 
of ACS, the main body of the galaxy was contained on WF1. Hence 
we could use the star counts on the other CCD to 
firstly check the predicted star counts and secondly to help identify
the bin with the statistical excess in $V$. The galaxy disk and halo 
may well extend onto the WF2 chip, but it served as a useful check. 
We wished to identify the
brightest supergiants in $V$ using the field stars, both because the calibrations for RSGs are in $V$ and \citet{1981ApJS...47..357B}
do not have predictions for stars in this band. Both the predicted
star counts and the number of field stars were scaled to the
field-of-view of the galaxy data used
(149\arcsec$\times$92\arcsec). The luminosity functions for red and blue supergiants are given in Tables~\ref{tab:RSGs} and~\ref{tab:BSGs}, and plotted in Figs.~\ref{fig:Rlfn} and
\ref{fig:Blfn}, respectively.

\begin{table}
  \caption{{\it VI}-band luminosity functions of RSGs in NGC~6207, stars in a $149''\times92''$ field adjacent to NGC~6207 and predicted foreground stars from \citet{1981ApJS...47..357B} field~16.}\label{tab:RSGs}
  \centering
  \begin{tabular}{lrrr} 
    \hline
    $V$ &  $n_{\rm RSG}$ & $n_{\rm field}$ & $n_{\rm fg}$\\
    \hline
    18--19 & 0      &  0	&  -- \\
    19--20 & 0      &  0	&  -- \\
    20--21 & 3      &  2	&  -- \\
    21--22 & 2      &  1	&  -- \\
    22--23 & 4      &  3	&  -- \\
    23--24 & 35     &  5	&  -- \\
    24--25 & 116    &  7	&  -- \\
    25--26 & 786    &  19	&  -- \\
    26--27 & 3066   &  98	&  -- \\
    \hline
    \hline
    $I$ &  $n_{\rm RSG}$ & $n_{\rm field}$ & $n_{\rm fg}$\\
    \hline
    18--19 & 1      & 0      & 1.50 \\
    19--20 & 5      & 3      & 2.02 \\
    20--21 & 2      & 4      & 2.60 \\
    21--22 & 7      & 4      & 3.20 \\
    22--23 & 57     & 4      & 3.71 \\
    23--24 & 442    & 7      & 3.63 \\
    24--25 & 1453   & 16     & 4.30 \\
    25--26 & 1921   & 55     & 4.60 \\
    26--27 & 128    & 36     & --   \\
    \hline 
  \end{tabular}
\end{table}

\begin{table}
  \caption{$B$-band luminosity functions of BSGs in NGC~6207, stars in a $149''\times92''$ field adjacent to NGC~6207 and predicted foreground stars from \citet{1981ApJS...47..357B} field~16.}\label{tab:BSGs}
  \centering
  \begin{tabular}{lrrr}
    \hline
    $B$  &  $n_{\rm BSG}$ & $n_{\rm field}$ &   $n_{\rm fg}$\\
    \hline
    18--19  & 0      &  0  &  0.50 \\
    19--20  & 0      &  0  &  0.68 \\
    20--21  & 0      &  1  &  0.89 \\
    21--22  & 1      &  1  &  1.12 \\
    22--23  & 12     &  0  &  1.36 \\
    23--24  & 114    &  5  &  1.61 \\
    24--25  & 679    &  2  &  1.87 \\
    25--26  & 2656   &  9  &  2.14 \\
    26--27  & 4961   &  41 &  --   \\
    \hline
  \end{tabular}
\end{table}

The predicted star counts and the number of field stars are in good agreement for magnitudes less than 23. A significant statistical excess of RSGs was found in the range $I = 22\!-\!23$ and BSGs in the range in $B = 22\!-\!23$. The three RSGs that were brightest in $V$, in the centre of the range $I = 22\!-\!23$, have an average of $V = 23.10$. This agrees well with the bin in the $V$-band luminosity function, which appears to have a statistical excess when compared to the field star count.  We also fitted a linear function to plots of $I$ against $V$, to convert our $I$ magnitude to $V$. The results were consistent with the $V$-band luminosity function, so the bin $V = 23\!-\!24$ was taken to contain the brightest RSGs in $V$. The brightest RSGs and BSGs were therefore taken to have $V = 23.5\pm0.5$ and $B = 22.5\pm0.5$, respectively.

We firstly calculated the distance using the calibrations of \citet{1994MNRAS.271..530R} with \mbox{$B_0^T = 11.16$}, which is the face-on total apparent magnitude of NGC~6207 corrected for Galactic extinction, and for the inclination of NGC~6207, from the Lyon-Meudon Extragalactic Database\footnote{http://leda.univ-lyon1.fr}~(LEDA). We used an iterative approach with the calibrations from \citet{1994MNRAS.271..530R} using $M^{\rm gal}_B = -19.74$, which has been corrected to $H_0 = 72$~\hub, also from LEDA. The results are listed in Table~\ref{tab:roz}, where 10(a) and (d) are for RSGs and 10(c) and (f) are for BSGs. The errors on the distance moduli, and the corresponding distances, are a combination of the errors associated with the regressions of \citet[table 5]{1994MNRAS.271..530R}, and errors in quadrature for the size of the bin and the reddening. The simple averages give a distance modulus of $\mu =31.64^{+ 0.33}_{-0.40}$, and distance of $D=21.40^{+ 3.38}_{-3.66}$\,Mpc, where the errors reflect the range of the individual values calculated.

\begin{table*}
\caption{Results of the brightest supergiants distance estimate for NGC~6207 using \citet{1994MNRAS.271..530R}. See Section~\ref{sec:BSGs} for details.}\label{tab:roz}
  \centering
  \begin{tabular}{lrrrr}\hline
     & 10(a) & 10(c) & 10(d) & 10(f)\\
    \hline
    $\mu$ & 31.97(0.77) & 31.70(1.03) & 31.63(0.74) & 31.24(1.01)\\
    $D$ (Mpc) & 24.78(8.79) & 21.89(10.38) & 21.19(7.22) & 17.74(8.25)\\
    \hline
  \end{tabular}
\end{table*}

The calibrations given in equations~(1) and~(2) of
\citet{1994A&A...286..718K} are of the same form as the regressions
10(a) and (c) of \citet{1994MNRAS.271..530R}. We used the same
iterative approach here, although we used $B_0^T = 11.74$, which is
not corrected for the galaxy's inclination, to be consistent with the
data used to find the relationships. The distance moduli found were
$\mu(V) = 31.84\pm0.59$ and $\mu(B) = 32.14\pm0.59$, corresponding to
$D(V) = 23.32\pm6.34$ and $D(B) = 26.79\pm7.28$\,Mpc. The errors are a
combination of the errors associated with the regressions, the size of
the bin and the reddening. The averages give $\mu = 31.99\pm0.15$ and
$D = 25.06\pm1.74$\,Mpc, where the errors again reflect the range of the individual values calculated.

\subsection{Summary of distance estimates }\label{sec:D}

\begin{table}
\caption{Summary of distance estimates to NGC~6207. The estimated uncertainties are
given in parentheses.}
  \centering
  \begin{tabular}{lrrr}\hline
    \label{tab:D}
    Method & Source & Distance & Mean\\
    & & (Mpc)& \\
    \hline
    Standard Candle       & 1 & 21.0(4.3) & 21.0(4.3)\\[4pt]
    Brightest supergiants & 1 & 21.4(3.5)& 23.2(4.5)\\
                          & 1 & 25.1(1.7)&\\[4pt]
    Kinematic             & 2 & 18.1 & 16.6(3.3)\\
    & 3 & 15.1 &\\[5pt]
    Mean & & & 20.3(3.4)\\
    \hline
  \end{tabular}
\flushleft (1)~this paper,
(2)~\citet{1988ngc..book.....T},
(3)~LEDA
\end{table}

A summary of distances to NGC~6207 from three different methods is listed in Table~\ref{tab:D}. As well our SCM and BSM estimates, derived in Sections~\ref{sec:SCM} and \ref{sec:BSGs}, we have collated two kinematic distance estimates from the literature. The distance from \citet{1988ngc..book.....T} uses a heliocentric velocity of 852\,\kms\ and the \citet{1984ApJ...281...31T} model for infall onto the Virgo cluster. The distance from LEDA uses a heliocentric velocity of 851\,\kms\ and instead uses the \citet{1998A&AS..130..333T} model for infall onto the Virgo cluster. The values have been corrected for a Hubble constant of $H_0 = 72$~\hub, in keeping with the rest of this paper. The simple mean of these kinematic distances is $D = 16.6\pm3.3$\,Mpc, where the error is a combination of the standard deviation of the distances and the uncertainty resulting from the cosmic thermal velocity dispersion of 187\,\kms\ \citep{2000ApJ...530..625T}. The accuracy of kinematic distances is limited not only by the observed velocity dispersion around the Hubble Flow, but on an accurate model to account for the infall onto Virgo and the Great Attractor, hence there is a large uncertainty associated with this method. Table~\ref{tab:D} also gives the simple mean of the BSM estimate, where the error reflects the range of individual values estimated.

The BSM distance estimate yielded fairly consistent results for NGC~6207, although they had large errors associated with them. The individual estimates range from 17.7--26.8\,Mpc, spanning 9.1\,Mpc, over both the calibrations from \citet{1994MNRAS.271..530R} and \citet{1994A&A...286..718K}. The standard deviations for each of these calibrations were 2.9 and 2.5\,Mpc respectively, which are high compared to the standard deviations that are usual for this method. \citet{1994MNRAS.271..530R} tested this method for several galaxies, with distances less than 5\,Mpc, using the same four relationships as we have used here. They found standard deviations of around 0.4\,Mpc with the exception of 1.7\,Mpc for NGC~4395. In addition to the tests of \citet{1994MNRAS.271..530R}, M74 was found to be at a distance of 7.7\,Mpc with a standard deviation of 0.8\,Mpc using these calibrations \citep[see][]{2005MNRAS.359..906H}. The data used to produce these calibrations were all below $\sim$7\,Mpc, therefore there could be a problem with extrapolating this relationship to higher distances. The accuracy of this method decreases as the distance increases because of contamination by unresolved clusters, and the calibrations themselves were found using data with distances less than $\sim$7\,Mpc.

The distance estimates range from 15.1--25.1\,Mpc, which spans 10\,Mpc over all the methods, illustrating the inherent problems with distance estimation. All of these methods have problems and large errors associated with them. A full, in depth discussion of these problems can be found in \citet{2005MNRAS.359..906H}. In order to obtain a distance, which is not biassed towards any one method, we have taken a simple mean of the average results from each method to find the distance to NGC~6207. We find the distance to NGC~6207 to be $D = 20.3\pm3.4$\,Mpc, where the error is the standard deviation of the averages of the individual methods.

\section{Estimate of ejected $^{56}$Ni mass}\label{sec:Ni}

The light curves of SN~2004A and SN~1999em in Fig.~\ref{fig:04A99em}
are very similar, although the
tail of SN~2004A may be slightly brighter. We would therefore expect the nickel
mass to be comparable to $\mni = 0.048$\,\msun, the nickel mass of
SN~1999em from \citet{2003ApJ...582..905H}, which is corrected for the
revised distance from \citet{2003ApJ...594..247L}, as discussed in 
\citet{2005MNRAS.359..906H}. We have used the
same three methods as \citet{2005MNRAS.359..906H} to find the mass of
$^{56}$Ni produced by SN~2004A. First of all it was estimated using
the bolometric luminosity of the exponential tail
\citep{2003ApJ...582..905H}, secondly using a direct comparison with
the light curve of SN~1987A and lastly using the `steepness of
decline' correlation, a new though unconfirmed method
\citep{2003A&A...404.1077E}.

\subsection{Nickel mass from bolometric luminosity of exponential tail}

\citet{2003ApJ...582..905H} derives $M_{\rm{Ni}}$ from the bolometric luminosity of the exponential tail, assuming that all of the $\gamma$-rays resulting from the $^{56}\rm{Co}\rightarrow ^{56}$Fe decay are fully thermalised. We first converted our $V$-band photometry in the tail to bolometric luminosities using equation~(1) of \citet{2003ApJ...582..905H}, given here in equation~(\ref{equ:bol}). The bolometric correction is $BC = 0.26\pm0.06$ and the additive constant converts from Vega magnitudes to cgs units \citep{2001PhDT,2003ApJ...582..905H}. We have used the reddening and distance derived in Sections~\ref{sec:red} and \ref{sec:D}, respectively.
\begin{equation}
\log \left(\frac{L}{{\rm erg\:s}^{-1}}\right) = \frac{-(V-A_V+ BC)+5\log D-8.14}{2.5}\label{equ:bol}
\end{equation}
\begin{equation}
M_{{\rm Ni}}= 7.866\times10^{-44} L \exp \left[ \frac{(t-t_0)/(1+z)-\tau_{{\rm Ni}}} {\tau_{\rm Co}} \right]\,\msun \label{equ:MNi}
\end{equation}
The nickel mass was then found using equation~(2) of \citet{2003ApJ...582..905H}, given here in equation~(\ref{equ:MNi}), where $t_0$ is the explosion epoch, $\tau_{{\rm Ni}} = 6.1$\,d is the half-life of $^{56}$Ni and $\tau_{{\rm Co}} = 111.26$\,d is the half-life of $^{56}$Co. We have used the explosion epoch derived in Section~\ref{sec:epoch} and the redshift is from NED.

Using this method we estimated \mni\ for each of the three points in the tail. We further estimated the range of possible values that this mass can possibly obtain, using the errors for each parameter. A simple average of the \mni\ results gives $\mni = 0.050^{+0.040}_{-0.020}$, which is comparable to the nickel mass of SN~1999em as we expected. The error was calculated from the average of the extremes of the nickel mass values and not from the standard deviation, which was 0.013\,\msun, as the standard deviation only reflects the scatter in $V$-band magnitudes in the tail. This method is strongly dependent on the explosion epoch, extinction and the distance. Although the explosion epoch is fairly well known, there is a large uncertainty associated with distance and the $V$-band photometry, hence the large error.

\subsection{Nickel mass from a direct comparison to SN~1987A light curve} 

The nickel mass was also estimated from the difference in the pseudo-bolometric (UVOIR) light curves of SN~2004A and SN~1987A, assuming the same $\gamma$-ray deposition. A $\chi^2$-fitting algorithm was used to shift the light curve of SN~1987A onto that of SN~2004A to find the best fit. When constructing the light curve of SN~1987A a distance of 50\,kpc was adopted. The difference in log luminosity was found to be $\log (L^{87A}/L) = 0.247 \pm 0.030$. Equation~(\ref{equ:Ni87A}) was then used to scale the nickel mass of SN~1987A, which was taken to be 0.075\,\msun\ \citep[e.g.][]{1998ApJ...498L.129T}, to estimate that of SN~2004A.
\begin{equation}
M_{\rm Ni} = 0.075 \times \left(\frac{L}{L^{87A}}\right)\,\msun
\label{equ:Ni87A}
\end{equation}
In this way we find that $\mni = 0.042^{+0.017}_{-0.013}$\,\msun, where the error is the combined error from the fit, the uncertainty in the explosion date and the distance. This method is dependent on two assumptions, one is the distance to the SN and the error in the nickel mass amply accounts for the distance uncertainty. The second is the assumption that both SNe have similar $\gamma$-ray escape fractions deposition rates. The validity of this cannot be tested with the current data, and this assumption is likely to lead to larger errors than the distance uncertainty. 

\subsection{Nickel mass from `steepness of decline' correlation} 

\citet{2003A&A...404.1077E} reported a correlation between the rate of decline in the $V$ band, from the plateau to the tail, and the nickel mass estimated from the SN~1987A method. The advantage with this method is that it is independent of distance and explosion epoch, however is does require a reasonably well populated light curve and at the moment is an unconfirmed method. The authors defined a `steepness' parameter, $S$, which is the maximum gradient during the transition in mag\,d$^{-1}$. A sample of ten SNe~II-P were used to determine the best linear fit. We have re-examined the results of \citet{2003A&A...404.1077E} and derive the relationship given in equation~(\ref{equ:S}).
\begin{equation}
\log \left(\frac{\mni}{\msun}\right) =  -6.9935(\pm 0.3791)S-0.7383(\pm0.0355)
\label{equ:S}
\end{equation}
Unfortunately the light curve of SN~2004A is quite sparse so we could only put limits on the nickel mass using this method. We fitted a series of straight lines to the data to find the maximum and minimum values of S, which were plausible. The `steepness' parameter for SN~2004A, using this method, was found to be in the range 0.07--0.21, which corresponds to $\mni = 0.006-0.056$\,\msun, which is consistent with the first two methods.

\subsection{Discussion of nickel mass estimate}\label{sec:concni}

The \citet{2003ApJ...582..905H} and the direct comparison methods give consistent results of $\mni = 0.050^{+0.040}_{-0.020}$ and $0.042^{+0.017}_{-0.013}$\,\msun, which are comparable to the nickel mass of SN~1999em, as we would expect from the appearance of the light curve and the spectra. These methods are however both dependent on the distance, although the errors sufficiently accommodate this. The method of \citet{2003ApJ...582..905H} assumes that the $\gamma$-rays are fully thermalised, but this is not unreasonable as the slope of the exponential tail of SN~2004A roughly follows that of SN~1999em, who's rate of decay in the tail phase followed the decay of $^{56}{\rm Co}$ \citep{2003MNRAS.338..939E}. The direct comparison to the light curve of SN~1987A assumes that SN~2004A deposited a similar fraction of $\gamma$-rays to SN~1987A. As the late-time decline of SN~1987A was also very close to the decay rate of $^{56}$Co, this assumption is also not unfounded.

It is unfortunate that the light curve of SN~2004A was too sparse to obtain a reliable distance independent result from the `steepness' method. From the limits that were
obtained using this method, the nickel mass is unlikely to be as low
as 0.006\,\msun. This is the same nickel mass, from
\citet{2004MNRAS.347...74P} corrected to the revised distance of
\citet{2005MNRAS.359..906H}, as the low nickel, low luminosity
(`faint') SN~1997D. SNe with low nickel masses
\citep[e.g.][]{2003MNRAS.338..711Z,astro-ph/0310056,2004MNRAS.347...74P}
have low luminosities and velocities, hence have narrow features in
their spectra. SN~2004A has none of these features and appears to be
more similar to SN~1999em. 
The upper limit from the `steepness of decline' method is consistent with the
\citet{2003ApJ...582..905H} and the direct comparison with SN~1987A
methods. We therefore take the average of the former two methods to get
$\mni = 0.046^{+0.031}_{-0.017}$, where the error is the combined
error of the methods.


\section{Analysis of the progenitor images}\label{sec:prog}

\subsection{Metallicity of NGC~6207 at the position of SN~2004}\label{sec:Z}
As can be seen in Fig.\,\ref{fig:finder} the SN occurred at 
a significant distance from the centre of the galaxy NGC~6207. 
Given that the host galaxy is an Sc, with a likely strong abundance 
gradient \citep[e.g.][]{1992MNRAS.259..121V}, it is  
possible that the environment was significantly metal poor and hence
the progenitor star may have been of lower than solar metallicity. 
There is no measurement of the abundance gradient in NGC~6207, but 
as an illustrative argument we can estimate the likely metallicity at the
galactic position of SN~2004A from the properties of NGC~6207. At a distance
of 20.3$\pm$3.4\,Mpc, NGC~6207 has 
$M_{\rm B}=-20.0\pm0.3$ from LEDA.
Using the correlations between oxygen abundances and macroscopic galactic
properties of \citet{2004A&A...425..849P}, NGC~6207 should have  
an oxygen abundance at $0.4R_{25}$ of 
$12 + \log {\rm O/H} \simeq 8.5 \pm 0.2$. 
Again using the parameters from LEDA, we find that $R_{25}$=4\,kpc
and that SN~2004A lies at a galactocentric distance of $R_{g}=6.2$\,kpc. 
The latter number assumes that SN~2004A occurred in the disk of NGC~6207 and
is a deprojected distance assuming a disk inclination angle of 
68.4$^{\circ}$ and a major axis angle of 17.5$^{\circ}$ (east of north). 
NGC~6207 appears to be an Sc galaxy, similar in morphology, size and absolute 
magnitude to M33 hence if we assume
that the stellar oxygen abundance gradient in M33 
\citep[$-0.05$\,dex\,kpc$^{-1}$;][]{Urbaneja_m33_gradient} is applicable in  
NGC~6207 then the oxygen abundance is likely to be $8.3\pm0.4$, where the 
error comes from a combination of the error estimate at $0.4R_{25}$ and 
the typical scatter in abundance gradients of Sc galaxies of 
$\pm0.05$\,dex\,kpc$^{-1}$. Although the error is large, it suggests that 
the metallicity of the progenitor star was likely a factor of two 
below solar, similar to the Large Magellanic Cloud. The spectrum of
SN~2004A does not show any distinct peculiarities compared to that of 
SN~1999em which likely arose from a RSG of approximately 
solar metallicity \citep{2002ApJ...565.1089S}. It would be 
of interest to determine the metallicity accurately in the region 
where SN~2004A exploded, from the nearest accessible H\,{\sc ii} regions.

\subsection{Astrometry of {\it HST} observations}

To determine the position of SN~2004A on the pre-explosion images, we
employed the same differential method as in
\citet{2004Sci...303..499S} and \citet{2005MNRAS.360..288M}. The
position of SN~2004A was found to lie on the WF3 chip in all the
pre-explosion observations made using WFPC2. The F606W and F814W
pre-explosion observations were taken consecutively, and were found to
be aligned to better than 0.01 arcsec.  The positions of 16 stars,
common to the pre-explosion WF3 F814W and post-explosion ACS F555W
images, were measured using aperture photometry within the {\sc
daophot} package. The post-explosion F555W frame was transformed to
the WF3 pre-explosion co-ordinate system using the {\sc iraf} task
{\sc geomap}. The F300W pre-explosion image however was taken
$\sim$1\,year previously with a different telescope orientation, 
and a similar process was followed for alignment. 
Due to the pointing of the observation and
the short wavelength band pass of the filter, only three stars could
be identified on WF3 for use in the transformation calculation. It
was possible to estimate an approximate position for the SN to
roughly 0.14 arcsec, and since no objects are detected anywhere near
the SN position the exact position is not critical.

\begin{figure}
  \vspace{5mm}
  \centering
  \begin{minipage}[c]{0.5\textwidth}
    \centering
    \epsfig{file = 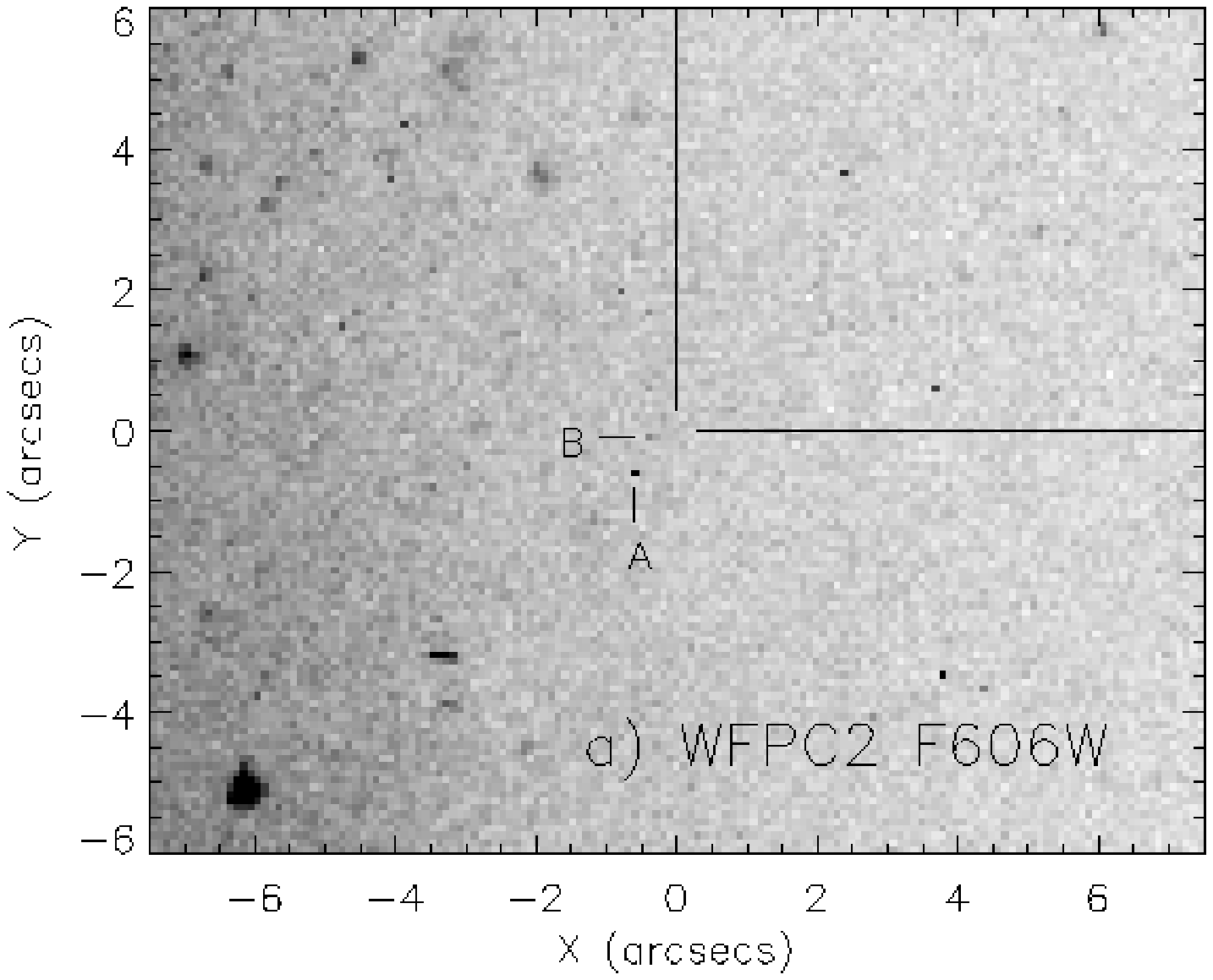, width = 90mm}
  \end{minipage}\\[10pt]
  \begin{minipage}[c]{0.5\textwidth}
    \centering
    \epsfig{file = 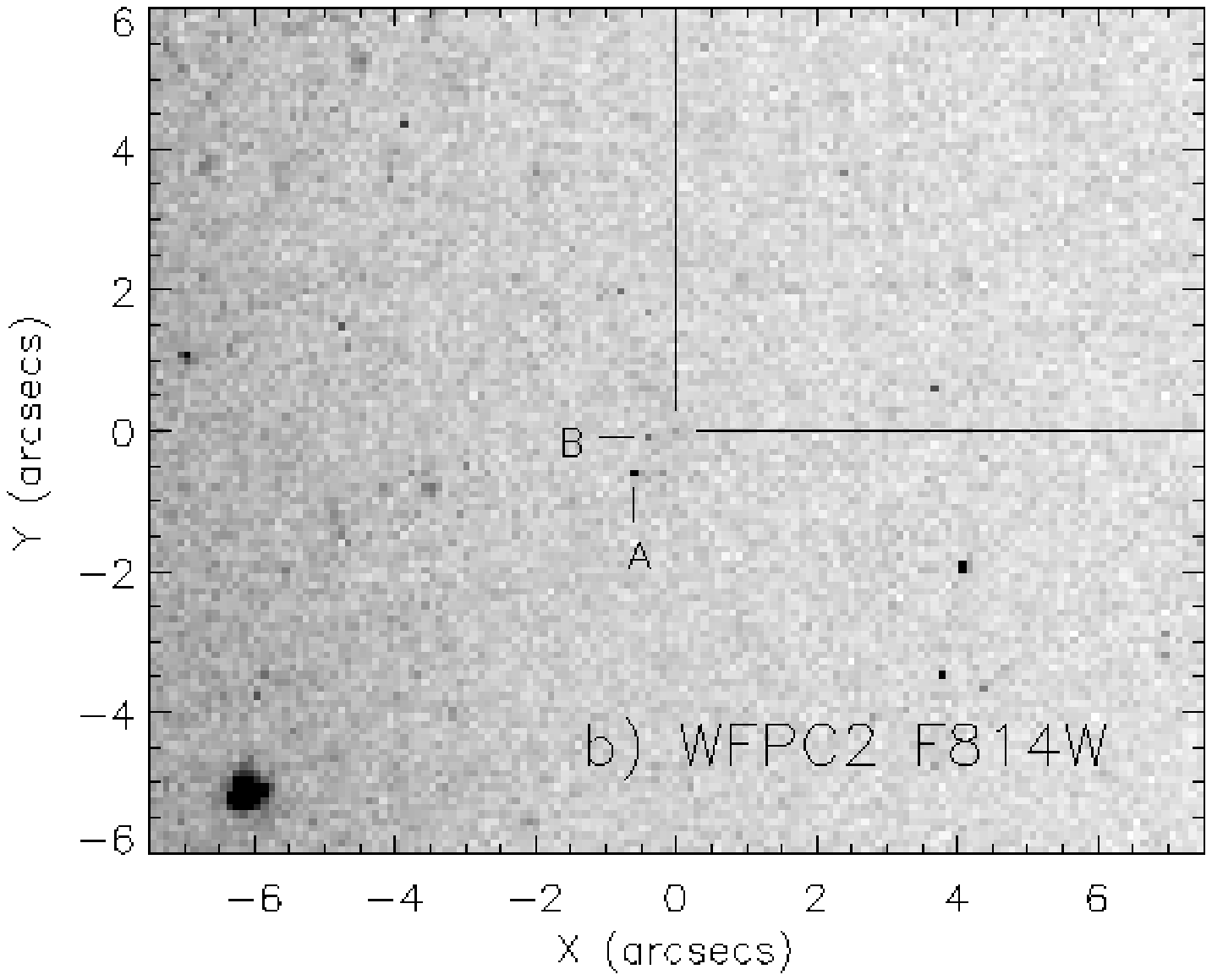, width = 90mm}
  \end{minipage}
  \begin{minipage}[c]{0.5\textwidth}
    \centering
    \epsfig{file = 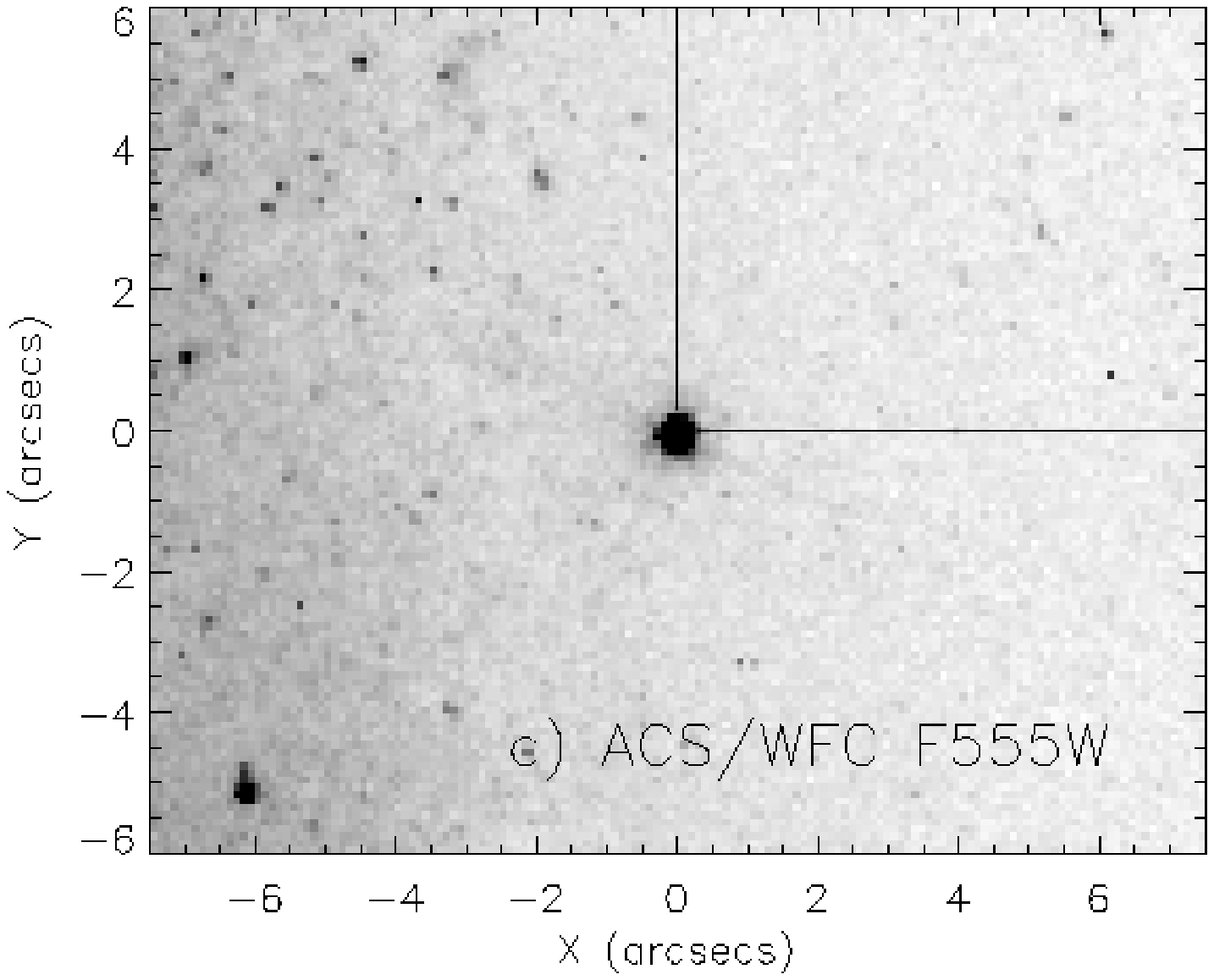, width = 90mm}
  \end{minipage}
  \caption{Pre- and post-explosion imaging of the site of SN 2004A.  All images are centred on the SN location. a) WFPC2 F606W pre-explosion image. b) WFPC2 F814W pre-explosion image.  The objects labelled A and B in these images are not stars but hot pixels. c) ACS/WFC F555W post-explosion image.  In the F814W pre-explosion image a faint object is just visible at the SN location.} \label{fig:prepost}
\end{figure}

The pixel location of SN~2004A was measured on the post-explosion frame, again using aperture photometry, and the same transformation was applied to this position to determine the coordinates of the SN in the pre-explosion images. In order to give an average value and estimated error, three different centring algorithms were employed to determine the SN centroid. Fig.~\ref{fig:prepost} shows sections of the pre- and post-explosion imaging centred on the SN coordinates.

\subsection{Detection of a progenitor star in F814W}

There is no progenitor object visible in either the F606W or the F300W
frames at the position of SN~2004A, however there is a faint object
detected in the F814W frames. The object has a significance of
4.7$\sigma$ measured with simple aperture photometry in a 2 pixel
aperture. The PSF-fitting WFPC2 photometry package {\sc hstphot}
\citep{2000PASP..112.1383D} also independently detects an object at
4.8$\sigma$, suggesting a star-like PSF. After aperture and CTE corrections, aperture photometry determined
a flight system magnitude of $m_{\rm F814W} = 24.3 \pm 0.3$. The {\sc
hstphot} magnitude was $m_{\rm F814W} = 24.4 \pm 0.2$, again after
similar corrections. The astrometric errors and the difference in the
position of the progenitor object and SN are listed in Table~\ref{tab:errors}. The error in the position of the progenitor star in
the under-sampled WFPC2 image was estimated from the range of four
methods used to determine the position i.e. three different centring
algorithms within the aperture photometry routines of {\sc daophot}
and the PSF-fitting method within {\sc hstphot}. The error in the SN
position was determined in a similar way. The geometric transformation
error is a combination of the RMS residuals from the 2-dimensional
spatial transformation functions. The total error of the differential
astrometric solution is calculated by combining these three
independent errors in quadrature. The difference in the position of
the SN is 34 milli-arcsec (mas), and the error in the method is 38
mas. Hence, within the errors, the progenitor object detected in the
F814W image is coincident with the SN position.

It is therefore probable that this object is the progenitor star of SN
2004A. The agreement between the two different methods, the flux (and
hence s/n), centroid, magnitude and error suggests that this is indeed
a significant detection, with a magnitude of $m_{\rm F814W} = 24.35
\pm 0.3$. Fig.~\ref{fig:prepost} shows that there are two hot pixels
near the position of SN~2004A, which are clearly labelled in the
associated data quality files released as part of the WFPC2 data
package. However there is no hot pixel flagged at the position of
SN~2004A in the data quality files that would suggest a spurious
detection. We have estimated 5$\sigma$ limiting magnitudes for each of
the filters, in the vicinity of the position of SN~2004A
(i.e. using an appropriate local sky flux), and these are $m_{F606W} =
25.4 \pm 0.3$, $m_{F814W} = 24.3 \pm 0.3$, $m_{F300W} = 23.1 \pm
0.3$. As expected the 5$\sigma$ limiting magnitude calculated for the
F814W filter is consistent with the estimated magnitude of the
4.7$\sigma$ object. In the discussion in Section~\ref{sec:dissprog},
we argue that whether or not we interpret this as a real detection of
the progenitor, or as an upper limit to the magnitude, the
implications for the progenitor star are similar.

\begin{figure}
  \vspace{5mm}
  \centering
  \begin{minipage}[c]{0.5\textwidth}
    \centering
    \epsfig{file = 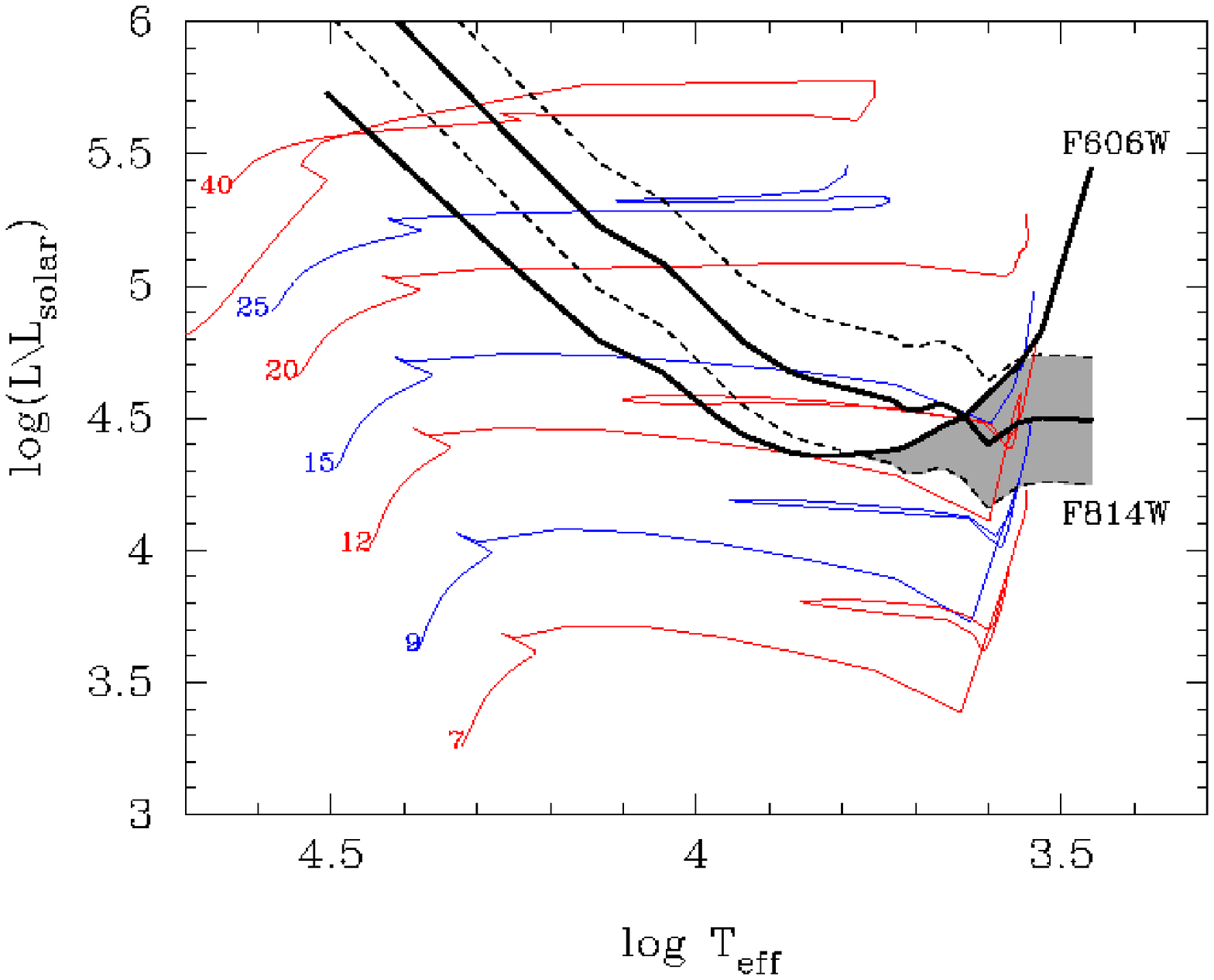, width = 90mm}
  \end{minipage}\\[10pt]
  \begin{minipage}[c]{0.5\textwidth}
    \centering
    \epsfig{file = 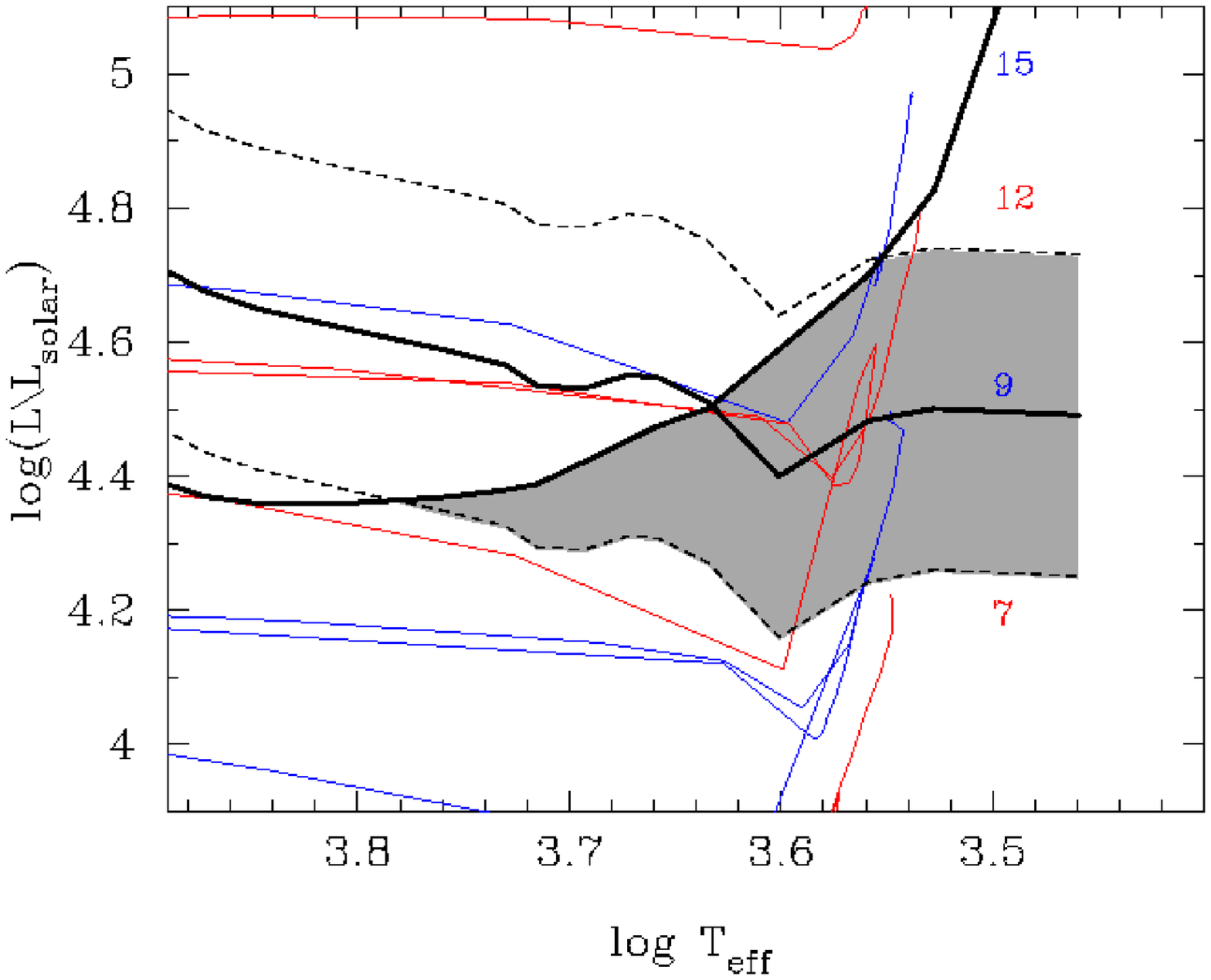, width = 90mm}
  \end{minipage}
  \caption{H--R diagram showing the Geneva stellar evolutionary tracks (coloured red and blue
alternatively for clarity, with initial masses labelled) and the likely position 
for the progenitor. The progenitor detected at 5$\sigma$ in the $I$-band has a 
luminosity dependent on effective temperature. This solid black curve, labelled F814W
is plotted and bracketed by the likely errors (dotted curve, see text). The solid 
black curve labelled F606W is the limiting luminosity implied by the sensitivity of the 
F606W pre-explosion images. All stars lying above this line would have been 
detectable in the diagram. Hence combining these two constraints 
suggests a region of possible existence for progenitor
stars. This region is shaded grey, and implies the progenitor was likely 
a RSGs with initial masses of $9^{+3}_{-2}$\,\msun. The lower panel is
a magnified view of the interesting shaded region shown in the upper panel. }
\label{fig:tracks}
\end{figure}

\subsection{Luminosity and mass limits for the progenitor of SN 2004A}\label{sec:dissprog}

First of all we assume that the 4.7$\sigma$ F814W object is a real detection of the progenitor star and combine this with the F606W limiting magnitude. The methods of \citet{2003MNRAS.343..735S} and \citet*{2005MNRAS.360..288M} were employed, in which the apparent flight system WFPC2 magnitudes are converted to bolometric luminosities appropriate for supergiant stars of a range of spectral types/effective temperatures.

The stellar evolutionary tracks used were the Geneva tracks \citep{1992A&AS...96..269S,1994A&AS..103...97M} for single stars with initial masses ranging from 7--40\,\msun\ and solar metallicity ($Z = 0.02$). The issue of the metallicity was discussed in Section~\ref{sec:Z}, and the evolutionary tracks of 8--12\,\msun\ stars are not particularly dependent on metallicity, unless extremely low values are reached, which is unlikely to be the case here. The limiting apparent magnitudes were converted firstly to absolute and then bolometric magnitudes. The distance to NGC~6207 was estimated as $20.3\pm 3.4$\,Mpc (see Section~\ref{sec:D}) and the foreground reddening in this direction is given as $\ebv = 0.015$, from NED. However this does not of course take into account the reddening effects caused by the gas and dust of the star's host galaxy. The reddening toward the bright supergiants in the immediate vicinity of the SN from the three-colour ACS images was estimated as, $A_U = 0.29 \pm 0.15$, $A_V = 0.19 \pm 0.09$, $A_I = 0.09 \pm 0.05$ in Section~\ref{sec:red}.

The apparent flight system magnitudes of the F814W and F606W
observations were converted to standard {\it VI} filter magnitudes using
the tabulated values in \citet{2005MNRAS.360..288M}. This correction
is dependent on the spectral type of the object and hence extinction
corrected absolute magnitudes appropriate for supergiant stars of
spectral types in the range of O9 to M5, (\teff\ $32000-2880$
K) were calculated from the $m_{\rm F606W}$ and $m_{\rm F814W}$ flight
system magnitudes. \citet{2001ApJ...556L..29S} have tabulated spectral
type dependent corrections for $m_{\rm F300W}$ magnitudes, $c_{\rm
V-300}$, which also include a colour correction to the standard $V$ band, necessary for conversion to bolometric magnitudes. These
absolute magnitudes were converted to bolometric magnitudes and
luminosities (Table~\ref{tab:mag}). The $I$-band luminosities with
errors were plotted on a H--R diagram (Fig.~\ref{fig:tracks}), thereby
defining a strip within which the object must be placed. To further
constrain progenitor parameters, the $V$-band 5$\sigma$ luminosity
limit was also plotted, and possible progenitors must lie below this
limit on the H--R diagram. The $I$-band detection was thereby
restricted to spectral types ranging from G5 to M5 ($\teff \sim
4900$ K to 2900 K), ruling out BSGs and suggesting that
the progenitor of SN~2004A was a RSG just prior to
explosion. The initial mass of the progenitor is estimated to be
$9^{+3}_{-2}$\,\msun. The value of 9\,\msun\ comes from the closest
estimated end point to the $I$-band magnitude, which is also
consistent with the F606W limit. The upper error comes from the
highest possible mass it is likely to have been, which is the highest mass
track that the observed limits overlap in any significant way, and the
lower limit is estimated in a similar way. This method was adopted in
\citet{2004Sci...303..499S}.

If the object detected in the F814W image is not a real detection, then the lines plotted on Fig.~\ref{fig:tracks} can both be considered upper limits and the mass of $9^{+3}_{-2}$\,\msun, becomes a very robust upper mass limit of $< 12$\,\msun. Hence the important point to arise from this analysis is that the progenitor object can have had an initial mass of no more than 12\,\msun, and was probably in the range $9^{+3}_{-2}$\,\msun.

\begin{table}
  \caption[]{Astrometric errors of the position of SN~2004A and the progenitor
    star detected in the F814W filter.}
  \begin{center}
    \begin{tabular}{lr} \hline
      Source of Error               &  Error (mas)\\
      \hline
      Position of progenitor         &  11\\
      Position of SN                 &  10\\
      Geometric transformation (RMS) &  35\\
      Total error                    &  38\\
      \\
      {\bf Measured difference}       &  {\bf 34}\\
      \hline
    \end{tabular}
  \end{center}
  \label{tab:errors}
\end{table}
  
\begin{table*}
  \caption[]{Luminosity limits on the progenitor of SN 2004A from the WFPC2 F300W ($U$), F606W ($V$) and F814W ($I$) pre-explosion observations. The luminosity values for the $I$-band are estimated from the 4.7$\sigma$ detection discussed above. The values for $U$ and $V$ are from the 5$\sigma$ limiting magnitudes. The corrections for the F300W magnitudes were taken from \citet{2001ApJ...556L..29S}.  Corrections for F606W and F814W taken from \citet*{2005MNRAS.360..288M}.  The error on all values of $\log(L/L_{\odot}) = 0.2$.}
  \begin{center}
    \begin{tabular}{lrrrr} \hline
      Spectral type & \teff & $\log L/L_{\odot}$ ($U$) & 
      $\log L/L_{\odot}$ ($V$) & $\log L/L_{\odot}$ ($I$) \\
      \hline
      O9  &  32000  &   $-$	&	5.73	&	6.27\\
      B0  &  28500  &   5.99	&	$-$	&	$-$ \\
      B2  &  17600  &   5.45	&	5.06	&	5.53\\
      B5  &  13600  &   5.53	&	4.79	&	5.23\\
      B8  &  11100  &   5.76	&	4.67	&	5.08\\
      A0  &   9980  &   $-$	&	4.56	&	4.96\\
      A2  &   9380  &   $-$	&	4.51	&	4.89\\
      A5  &   8610  &   $-$	&	4.43	&	4.79\\
      F0  &   7460  &   6.04	&	4.37	&	4.68\\
      F2  &   7030  &   $-$	&	4.36	&	4.65\\
      F5  &   6370  &   $-$	&	4.36	&	4.62\\
      F8  &   5750  &   $-$	&	4.37	&	4.59\\
      G0  &   5370  &   6.30	&	4.38	&	4.57\\
      G2  &   5190  &   $-$	&	4.39	&	4.53\\
      G5  &   4930  &   $-$	&	4.42	&	4.53\\
      G8  &   4700  &   $-$	&	4.45	&	4.55\\
      K0  &   4550  &   6.80	&	4.48	&	4.55\\
      K2  &   4310  &   $-$	&	4.50	&	4.51\\
      K5  &   3990  &   $-$	&	4.59	&	4.40\\
      M0  &   3620  &   7.52	&	4.70	&	4.48\\
      M2  &   3370  &   7.64	&	4.83	&	4.50\\
      M5  &   2880  &   8.38	&	5.44	&	4.49\\
      \hline
    \end{tabular}
  \end{center}
  \label{tab:mag}
\end{table*}

\section{Discussion}\label{sec:diss}

\subsection{What can we say about SN~2004A?}\label{sec:2004A?}

\begin{figure}
  \begin{center}
    \epsfig{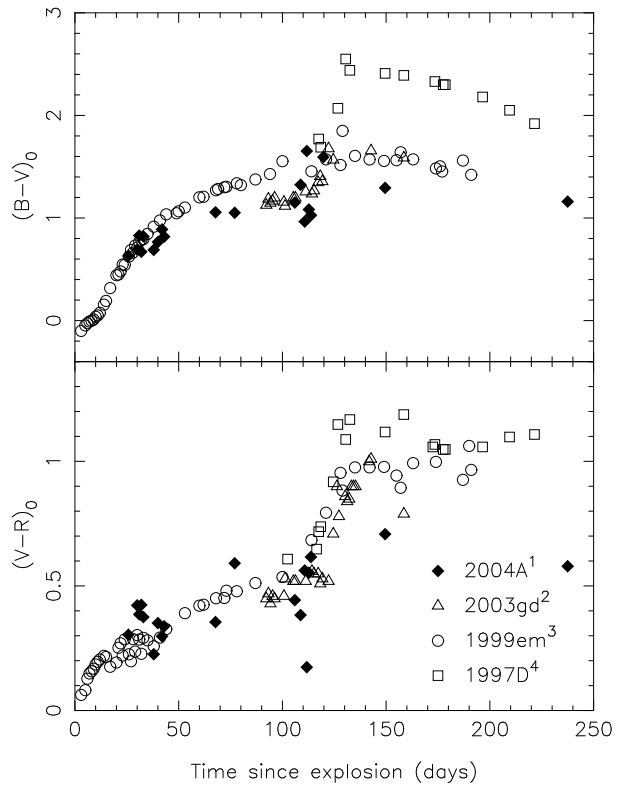}
    \caption{Intrinsic \bv\ and \vr\ colour curves of SN~2004A alongside those of the prototypical peculiar `faint' SN~1997D, the `normal' SNe~1999em and 2003gd. The superscripts in the figure denote the source of the photometry: (1) this paper, (2) \citet{2005MNRAS.359..906H}, (3) \citet{2001ApJ...558..615H} and (4) \citet{2001MNRAS.322..361B}.}\label{fig:diffSNc}
  \end{center}
\end{figure}

SN~2004A appears to be a `normal' SN II-P with a spectrum and light curve very like the well observed SN~1999em (Figs.~\ref{fig:s04A99em} and \ref{fig:04A99em}). The nickel mass found in Section \ref{sec:Ni} was also comparable to that of SN~1999em, as we would have expected. The existence of a low luminosity, low nickel mass, sub-group of SNe II-P \citep[e.g.][]{2003MNRAS.338..711Z,astro-ph/0310056,2004MNRAS.347...74P}, has been suggested. One of the proposed characteristics of this group is a rapid excess in the \bv\ and \vr\ colours at the end of the photospheric phase \citep{2004MNRAS.347...74P}. \citet{2003A&A...404.1077E} and \citet{2004MNRAS.347...74P} noted this rapid excess in the `faint' SN~1997D (see Fig.~\ref{fig:diffSNc}), and it was also observed by \citet{2004MNRAS.347...74P} in SN~1999eu, another `faint' SN, in the form of a sharp spike. Fig.~\ref{fig:c04A99em} compares the colour evolutions of SN~2004A and SN~1999em. The \bv\ colour evolution of SN~2004A clearly shows such a rapid excess at the end of the plateau, which is not seen in the colour curve of SN~1999em. Fig.~\ref{fig:diffSNc} shows a comparison of the colour curves of the prototypical peculiar `faint' SN~1997D \citep{2001MNRAS.322..361B}, and the `normal' SNe~1999em and 2003gd \citep{2001ApJ...558..615H,2005MNRAS.359..906H}. Although the excess is present in the \bv\ colour curve of SN~2004A, it does not show as great an excess as SN~1997D, but instead follows a similar curve to SN~2003gd with a slightly bluer tail. There is too large a scatter in the \vr\ colour curve to be of much use, but there does not appear to be any evidence of a colour excess. The appearance of the spectrum and the light curves imply that SN~2004A is a normal SN II-P, suggesting that this excess is not confined to the ranks of the `faint' SNe. Until we have a larger sample of SNe II-P, both `faint' and `normal', it is difficult to confirm or rule out this as a characteristic.

\begin{figure}
  \begin{center}
    \epsfig{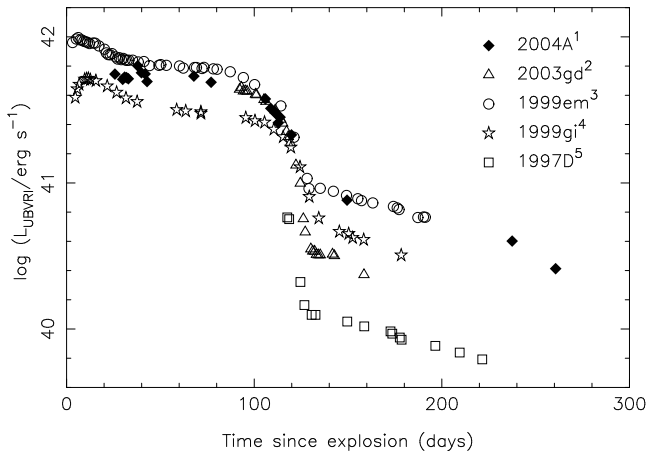}
    \caption{Comparison between the UVOIR light curves of SNe~2004A, 2003gd, 1999em, 1999gi and 1997D which have nickel masses of 0.046, 0.016, 0.048, 0.022 and 0.006\,\msun, respectively. The nickel mass for all the SNe, apart from SN~2004A which is given in this paper, are from \citet{2005MNRAS.359..906H}, and references therein. The superscripts in the figure denote the source of the photometry: (1) this paper, (2) \citet{2005MNRAS.359..906H}, (3) \citet{2001ApJ...558..615H}, (4) \citet{2002AJ....124.2490L} and (5) \citet{2001MNRAS.322..361B}.}\label{fig:UVOIR}
  \end{center}
\end{figure}

A comparison between the UVOIR light curves of SNe~2004A, 2003gd,
1999em, 1999gi and 1997D is shown in Fig.~\ref{fig:UVOIR}. All the
UVOIR light curves, apart from SN~2004A, are from
\citet{2005MNRAS.359..906H} and a full description of how they were
constructed can be found there. The UVOIR light curve of SN~2004A,
which was constructed in the same way, also confirms the similarities
between SN~2004A and the other normal SNe II-P.  SN~2004A differs slightly from SN~1999em in the plateau with SN~2004A being marginally fainter, whereas the tail luminosities reflect the comparable nickel mass ejected. 

\subsection{Implications for the progenitor of SN~2004A}

We compared the observed properties of SN~2004A with the progenitor
mass, presented in Section \ref{sec:prog}, using equation~(2) of
\citet{1985SvAL...11..145L}. The equations of
\citeauthor{1985SvAL...11..145L} relate the explosion energy, mass of
the envelope expelled (\mej) and initial radius of the star just
before the explosion, to the observed quantities of
SNe. \citet{2003MNRAS.346...97N} used a sample of 14 supernovae to test these
equations and found that they gave reasonable results. Equation~(2) of
\citet{1985SvAL...11..145L}, is shown here in equation~(\ref{equ:M}), where
\mej\ is the ejected envelope mass, $\Delta t$ is the length of the plateau and $u_{\rm ph}$ is the velocity of the photosphere material in the middle of the plateau.

\begin{eqnarray}
\label{equ:M}
\log \left(\frac{M_{\rm ej}}{\msun}\right) &=& 0.234 M_{\rm V}+2.91\log \left(\frac{\Delta t}{\rm days}\right)\nonumber\\
& + &1.96 \log \left(\frac{u_{\rm ph}}{10^3\; {\rm km\:s}^{-1}}\right)-1.829
\end{eqnarray}

If we consider the UVOIR light curve in Fig.~\ref{fig:UVOIR}, the
length of the plateau is $\sim$80\,d, although it is difficult to be
confident as the SN was not observed close to explosion. We therefore
estimate the length of the plateau to be $\Delta t = 80^{+25}_{-5}$\,d,
where the positive error is very conservative. Using the reddening
estimated in Section~\ref{sec:red}, the parameters discussed in
Section~\ref{sec:SCM} for the SCM at 50\,d post explosion and the
distance estimated in Section~\ref{sec:D}, we found that the ejecta
mass was $\mej = 11^{+10}_{-4}$. If we assume that the compact
remnant plus other mass losses from the system is $2\pm1$\,\msun\, then
it suggests that $\ms = 13^{+10}_{-4}$\,\msun\ for SN~2004A, where the upper mass limit of 23\,\msun\ is a
hard upper limit as the error in the plateau is very
conservative. The lower limit of this is consistent with the progenitor mass,
$9^{+3}_{-2}$\,\msun, found from pre-explosion images in
Section~\ref{sec:prog}. This result may also suggest that equation~(\ref{equ:M}) overestimates the mass somewhat as the lower limit is only just consistent even though our lower limits are fairly well constrained.

\begin{figure}
  \begin{center}
    \epsfig{file = 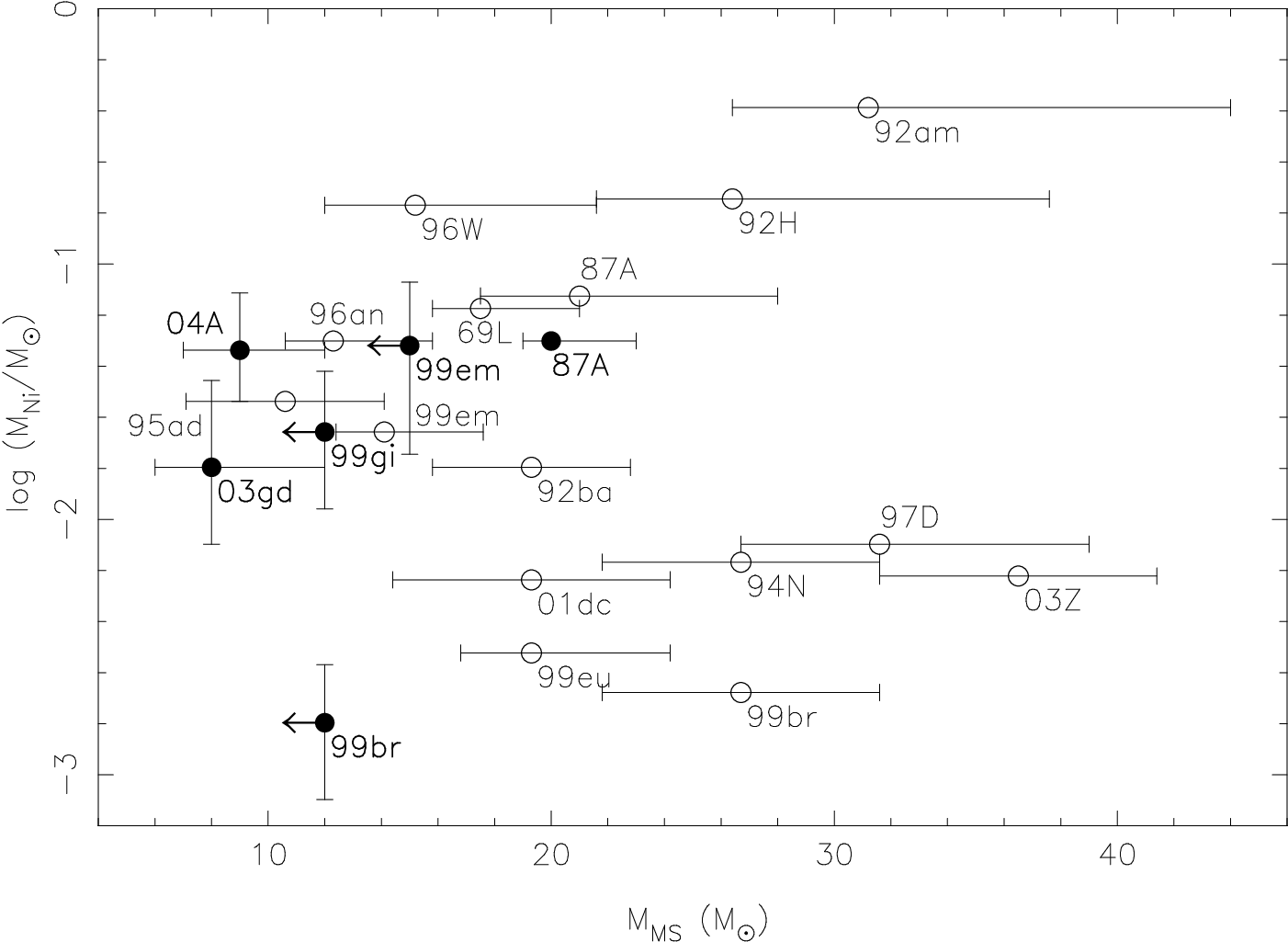,angle = -90,width = 80mm}
    \caption{Reproduction of fig.~1 (right) of \citet{astro-ph/0310057} showing the bimodal distribution of SNe II-P, in the \mni/\ms\ plane, from their model. The open circles show theoretical data from the semi-empirical model of \citet{2003MNRAS.338..711Z}, where \ms\ is estimated from the ejected envelope mass. The filled circles show observational data for SNe~2004A, 2003gd, 1999em and 1999br, where the observed nickel masses are estimated from the light curve. The nickel masses for SNe~2003gd and 1999em are from \citet{2005MNRAS.359..906H} and SN~1999br is from \citet{2003ApJ...582..905H}. \ms\ for SNe~1999em and 1999br are upper mass limits only and are from \citet{2002ApJ...565.1089S} and \citet{2005MNRAS.360..288M}, respectively. SN~2004A comfortably sits in the in the `normal' branch of the bimodal distribution.}\label{fig:zamp}
  \end{center}
\end{figure}

The observations of progenitors do not only have important
implications for stellar evolution theory, but also for the progenitor
models of `faint' SNe~II-P. There are two very different plausible
models for these, one being the low-energy explosion of massive stars
\citep[e.g.][]{1998ApJ...498L.129T,2001MNRAS.322..361B,2003MNRAS.338..711Z}. In
this model the collapsing core forms a black hole and a significant
amount of fallback of material occurs. An alternative scenario is the
low-energy explosion of low-mass stars, presented by
\citet{2000A&A...354..557C} who successfully reproduced the
observations of SN~1997D with an explosion of $10^{50}$ ergs and an
ejected mass of 6\,\msun. The findings of \citet{astro-ph/0310057}
support the high-mass progenitor scenario. The authors find a bimodal
distribution of SNe in the nickel mass, \ms\ plane. A
reproduction of the authors' Fig.~1 (right) is shown here in
Fig.~\ref{fig:zamp}. We are beginning to populate this figure with 
direct measurements of the main-sequence mass of the progenitors, as
opposed to model dependent values. The most interesting feature of
this figure, and the most controversial, is the existence of the 
faint branch. The SNe which populate this branch
 are proposed to have high initial mass progenitors. 
The points with direct measurements, or robust limits, of progenitor
mass from pre-explosion imaging are highlighted with the filled circles. 
This figure however should be treated with
caution as the theoretical estimates of \ms\ from
\citet{astro-ph/0310057}, shown with open circles in the figure, will
most probably need to be revised using the value of the opacity from
\citet{2005MNRAS.359..906H}. 
The observed nickel masses are estimated from the
light curves, and are from \citet{2003ApJ...582..905H} for SN~1999br and from \citet[and references therein]{2005MNRAS.359..906H} for SNe~2003gd and 1999em. The \citeauthor{2003MNRAS.338..711Z} nickel masses are however an input parameter of the semi-analytical light curve code and are therefore theoretical. \ms\ for SN~2003gd is from \citet{2004Sci...303..499S}, whereas
SNe~1999em and 1999br are upper mass limits only and are from
\citet{2002ApJ...565.1089S} and \citet{2005MNRAS.360..288M},
respectively. SN~2004A comfortably sits  in the `normal' branch
of the bimodal distribution. Even though the mass limit of SN~1999br
rules out a high mass progenitor for this `faint' SN, it does not
necessarily rule out a high mass progenitor scenario for other `faint'
SNe. It does however suggest that for at least some `faint' SNe a low
mass progenitor is likely. With more upper mass limits or direct
detections of the progenitor stars of these SNe, we should be 
able to determine the validity of the 
high mass progenitor scenario. 

\section{Conclusion}\label{sec:con}

We presented photometric and spectroscopic data of the Type II-P
SN~2004A, comparing the {\it BVRI} light curves with those of the well
observed SN~1999em using a $\chi^2$-fitting algorithm. This analysis
allowed us to estimate an explosion epoch of JD $245\,3011^{+3}_{-10}$,
corresponding to a date of 2004 January 6. We estimated the extinction
of the SN as $\ebv=0.06\pm0.03$, using {\it HST} ACS photometry of
the neighbouring stars and confirmed the reddening using the SN's
colour evolution. The expansion velocity was measured from our only
spectrum, and was found to be comparable to the velocities of similar
SNe II-P. This enabled us to extrapolate the velocity evolution of
SN~2004A forwards 7 days to find the velocity at a phase of 50\,d, to be
used in the SCM distance estimate.

Three new distances to NGC~6207 were calculated using two different
methods, the Standard Candle Method
\citep[SCM;][]{2003ApJ...582..905H,2004mmu..sympE...2H,astro-ph/0309122}
and the Brightest Supergiants Method
\citep[BSM;][]{1994MNRAS.271..530R,1994A&A...286..718K}. Firstly,
using the extrapolated velocity and interpolated {\it VI} magnitudes, we
estimated a distance of $D = 21.0\pm4.3$\,Mpc with the SCM. We
then used {\it HST} (ACS HRC) photometry to estimate the distances of
$21.4\pm3.5$ and $25.1\pm 1.7$\,Mpc using two different BSM
calibrations. Using these three distances and other distances within
the literature we estimated an overall distance of $20.3\pm 3.4$\,Mpc
to NGC~6207. This distance allowed the nickel mass synthesised in the
explosion to be estimated, $\mni = 0.046^{+0.031}_{-0.017}$\,\msun,
comparable to that of SN~1999em.

The probable discovery of the progenitor of SN~2004A on pre-explosion
{\it HST} WFPC2 images was presented. The star that exploded was 
likely to have been a RSG with a mass of $9^{+3}_{-2}$\,\msun. 
If the 4.8$\sigma$ detection is not believed then the 5$\sigma$
upper limit in the F814W images implies a robust upper mass limit for
a RSG progenitor of 12\,\msun. 
This is only the seventh progenitor star of an unambiguous core-collapse
SN that has had a direct detection. The first two progenitors discovered
were those of SN~1987A, which was a BSG, and SN~1993J,
which arose in a massive interacting binary system. Neither of these
fitted in with the theory that RSGs explode to give Type II
SNe. However, it appears that the elusive RSG progenitors are
now being found, with the discovery of the progenitor of SN~2004A
being the third (after SNe 2003gd and 2005cs). 

The observations of SN~2004A were compared to those of other SNe II-P
and were found to be consistent with other normal SNe II-P, although
there is a small peak in the UVOIR light curve at around 40\,d, which
could just be an artifact of the photometry. The SN observations were
then compared to the mass of the progenitor using theoretical
relationships which relate the mass of the envelope expelled to the
observed quantities of SNe, and were found also to be consistent. We
conclude that SN~2004A was a normal Type II-P SNe that arose from the
core-collapse induced explosion of a red supergiant of mass
$9^{+3}_{-2}$\,\msun.

\section*{Acknowledgments}

Based on observations made with the NASA/ESA Hubble Space Telescope,
obtained from the data archive at the Space Telescope Science
Institute. STScI is operated by the Association of Universities for
Research in Astronomy, Inc., under NASA contract NAS 5-26555. The 
spectroscopic data presented were obtained at the W.M. Keck Observatory,
which is operated as a scientific partnership among the California
Institute of Technology, the University of California and the National
Aeronautics and Space Administration. The Observatory was made
possible by the generous financial support of the W.M. Keck
Foundation.  The authors wish to recognize and acknowledge the very
significant cultural role and reverence that the summit of Mauna Kea
has always had within the indigenous Hawaiian community.  We are most
fortunate to have the opportunity to conduct observations from this
mountain. Some data presented were taken with the Liverpool Telescope
at the Observatorio del Roque de Los Muchachos, La Palma Spain. 
 The authors would
like to express their thanks to the research staff at Caltech and
Palomar Observatory who made the P60 automation possible.  The initial
phase of the P60 automation project was funded by a grant from the
Caltech Endowment, with additional support for this work provided by
the NSF and NASA.  A.~Gal-Yam acknowledges support by NASA
through Hubble Fellowship grant \#HST-HF-01158.01-A awarded by STScI,
which is operated by AURA, Inc., for NASA, under contract NAS
5-26555. S. Smartt acknowledges funding from PPARC and the European 
Science Foundation in the forms of Advanced and EURYI fellowships. M. Hendry thanks PPARC and Queen's University, Belfast for their financial support.

\label{lastpage}

\begin{thebibliography}{}

\bibitem[\protect\citeauthoryear{{Aldering}, {Humphreys} \&
  {Richmond}}{{Aldering} et~al.}{1994}]{1994AJ....107..662A}
{Aldering} G.,  {Humphreys} R.~M.,    {Richmond} M.,  1994, \aj, 107, 662

\bibitem[\protect\citeauthoryear{{Bahcall} \& {Soneira}}{{Bahcall} \&
  {Soneira}}{1981}]{1981ApJS...47..357B}
{Bahcall} J.~N.,  {Soneira} R.~M.,  1981, \apjs, 47, 357

\bibitem[Baron et al.(1985)]{1985PhRvL..55..126B} Baron, E., Cooperstein, 
J., \& Kahana, S.\ 1985, Physical Review Letters, 55, 126 

\bibitem[\protect\citeauthoryear{{Baron}, {Branch}, {Hauschildt}, {Filippenko},
  {Kirshner}, {Challis}, {Jha}, {Chevalier} \& {et al.}}{{Baron}
  et~al.}{2000}]{2000ApJ...545..444B}
{Baron} E.,  {Branch} D.,  {Hauschildt} P.~H.,  {Filippenko} A.~V.,  {Kirshner}
  R.~P.,  {Challis} P.~M.,  {Jha} S.,  {Chevalier} R.,    {et al.} 2000, \apj,
  545, 444

\bibitem[\protect\citeauthoryear{{Benetti}, {Turatto}, {Balberg}, {Zampieri},
  {Shapiro}, {Cappellaro}, {Nomoto}, {Nakamura}, {Mazzali} \&
  {Patat}}{{Benetti} et~al.}{2001}]{2001MNRAS.322..361B}
{Benetti} S.,  {Turatto} M.,  {Balberg} S.,  {Zampieri} L.,  {Shapiro} S.~L.,
  {Cappellaro} E.,  {Nomoto} K.,  {Nakamura} T.,  {Mazzali} P.~A.,    {Patat}
  F.,  2001, \mnras, 322, 361

\bibitem[\protect\citeauthoryear{{Cardelli}, {Clayton} \& {Mathis}}{{Cardelli}
  et~al.}{1989}]{1989ApJ...345..245C}
{Cardelli} J.~A.,  {Clayton} G.~C.,    {Mathis} J.~S.,  1989, \apj, 345, 245

\bibitem[Chevalier(1976)]{1976ApJ...207..872C} Chevalier, R.~A.\ 1976, 
\apj, 207, 872 

\bibitem[\protect\citeauthoryear{{Chugai} \& {Utrobin}}{{Chugai} \&
  {Utrobin}}{2000}]{2000A&A...354..557C}
{Chugai} N.~N.,  {Utrobin} V.~P.,  2000, \aap, 354, 557

\bibitem[\protect\citeauthoryear{{Dolphin}}{{Dolphin}}{2000}]{2000PASP..112.1383D}
{Dolphin} A.~E.,  2000, \pasp, 112, 1383

\bibitem[\protect\citeauthoryear{{Drilling} \& {Landolt}}{{Drilling} \&
  {Landolt}}{2000}]{2000asqu.book.....DL}
{Drilling} J.~S.,  {Landolt} A.~U.,  2000, {in Allen's astrophysical
  quantities}.
4th ed. Publisher: New York: AIP Press; Springer, 2000. Edited by Arthur
  N.~Cox. ISBN: 0387987460

\bibitem[\protect\citeauthoryear{{Eastman}, {Schmidt} \& {Kirshner}}{{Eastman}
  et~al.}{1996}]{1996ApJ...466..911E}
{Eastman} R.~G.,  {Schmidt} B.~P.,    {Kirshner} R.,  1996, \apj, 466, 911

\bibitem[\protect\citeauthoryear{{Eldridge} \& {Tout}}{{Eldridge} \&
  {Tout}}{2004}]{2004MNRAS.353...87E}
{Eldridge} J.~J.,  {Tout} C.~A.,  2004, \mnras, 353, 87

\bibitem[\protect\citeauthoryear{{Elmhamdi}, {Chugai} \& {Danziger}}{{Elmhamdi}
  et~al.}{2003}]{2003A&A...404.1077E}
{Elmhamdi} A.,  {Chugai} N.~N.,    {Danziger} I.~J.,  2003, \aap, 404, 1077

\bibitem[\protect\citeauthoryear{{Elmhamdi}, {Danziger}, {Chugai},
  {Pastorello}, {Turatto}, {Cappellaro}, {Altavilla}, {Benetti}, {Patat} \&
  {Salvo}}{{Elmhamdi} et~al.}{2003}]{2003MNRAS.338..939E}
{Elmhamdi} A.,  {Danziger} I.~J.,  {Chugai} N.,  {Pastorello} A.,  {Turatto}
  M.,  {Cappellaro} E.,  {Altavilla} G.,  {Benetti} S.,  {Patat} F.,    {Salvo}
  M.,  2003, \mnras, 338, 939

\bibitem[\protect\citeauthoryear{{Freedman}, {Madore}, {Gibson}, {Ferrarese},
  {Kelson}, {Sakai}, {Mould}, {Kennicutt} \& {et al.}}{{Freedman}
  et~al.}{2001}]{2001ApJ...553...47F}
{Freedman} W.~L.,  {Madore} B.~F.,  {Gibson} B.~K.,  {Ferrarese} L.,  {Kelson}
  D.~D.,  {Sakai} S.,  {Mould} J.~R.,  {Kennicutt} R.~C.,    {et al.} 2001,
  \apj, 553, 47


\bibitem[Gal-Yam et al.(2004)]{2004AAS...205.4006G} Gal-Yam, A., Cenko, 
S.~B., Fox, D.~W., Leonard, D.~C., Moon, D.-S., Sand, D.~J., \& Soderberg, 
A.~M.\ 2004, American Astronomical Society Meeting Abstracts, 205, 

\bibitem[Gal-Yam et al.(2005)]{2005ApJ...630L..29G} Gal-Yam, A., et al.\ 
2005, \apjl, 630, L29 

\bibitem[\protect\citeauthoryear{{Hamuy}}{{Hamuy}}{2001}]{2001PhDT}
{Hamuy} M.,  2001, Ph.D.~Thesis

\bibitem[\protect\citeauthoryear{{Hamuy}}{{Hamuy}}{2003}]{2003ApJ...582..905H}
{Hamuy} M.,  2003, \apj, 582, 905

\bibitem[\protect\citeauthoryear{{Hamuy}}{{Hamuy}}{2004a}]{2004mmu..sympE...2H}
{Hamuy} M.,  2004a, in Measuring and Modeling the Universe, from the Carnegie
  Observatories Centennial Symposia. Carnegie Observatories Astrophysics
  Series. Edited by W. L. Freedman, 2004. Pasadena: Carnegie Observatories,
  {The Latest Version of the Standardized Candle Method for Type II Supernovae,
  astro-ph/0301281}

\bibitem[\protect\citeauthoryear{{Hamuy}}{{Hamuy}}{2004b}]{astro-ph/0309122}
{Hamuy} M.,  2004b, in Cosmic Explosions. On the 10th Anniversary of SN~1993J
  (IAU Colloquium 192). Springer, Heidelberg, p. 535 {The Standard Candle
  Method for Type II Supernovae and the Hubble Constant, astro-ph/0309122}

\bibitem[\protect\citeauthoryear{{Hamuy} \& {Pinto}}{{Hamuy} \&
  {Pinto}}{2002}]{2002ApJ...566L..63H}
{Hamuy} M.,  {Pinto} P.~A.,  2002, \apjl, 566, L63

\bibitem[\protect\citeauthoryear{{Hamuy}, {Pinto}, {Maza}, {Suntzeff},
  {Phillips}, {Eastman}, {Smith}, {Corbally} \& {et al.}}{{Hamuy}
  et~al.}{2001}]{2001ApJ...558..615H}
{Hamuy} M.,  {Pinto} P.~A.,  {Maza} J.,  {Suntzeff} N.~B.,  {Phillips} M.~M.,
  {Eastman} R.~G.,  {Smith} R.~C.,  {Corbally} C.~J.,    {et al.} 2001, \apj,
  558, 615

\bibitem[\protect\citeauthoryear{{Heger}, {Fryer}, {Woosley}, {Langer} \&
  {Hartmann}}{{Heger} et~al.}{2003}]{2003ApJ...591..288H}
{Heger} A.,  {Fryer} C.~L.,  {Woosley} S.~E.,  {Langer} N.,    {Hartmann}
  D.~H.,  2003, \apj, 591, 288

\bibitem[\protect\citeauthoryear{{Hendry}, {Smartt}, {Maund}, {Pastorello},
  {Zampieri}, {Benetti}, {Turatto}, {Cappellaro} \& {et al.}}{{Hendry}
  et~al.}{2005}]{2005MNRAS.359..906H}
{Hendry} M.~A.,  {Smartt} S.~J.,  {Maund} J.~R.,  {Pastorello} A.,  {Zampieri}
  L.,  {Benetti} S.,  {Turatto} M.,  {Cappellaro} E.,    {et al.} 2005, \mnras,
  359, 906

\bibitem[\protect\citeauthoryear{{Holtzman}, {Burrows}, {Casertano}, {Hester},
  {Trauger}, {Watson} \& {Worthey}}{{Holtzman}
  et~al.}{1995}]{1995PASP..107.1065H}
{Holtzman} J.~A.,  {Burrows} C.~J.,  {Casertano} S.,  {Hester} J.~J.,
  {Trauger} J.~T.,  {Watson} A.~M.,    {Worthey} G.,  1995, \pasp, 107, 1065

\bibitem[\protect\citeauthoryear{{Howarth}, {Murray}, {Mills} \&
  {Berry}}{{Howarth} et~al.}{2003}]{SUN50.24}
{Howarth} I.~D.,  {Murray} J.,  {Mills} D.,    {Berry} D.~S.,  2003, {Starlink
  User Note 50.24}.
Rutherford Appleton Laboratory

\bibitem[\protect\citeauthoryear{{Karachentsev} \& {Tikhonov}}{{Karachentsev}
  \& {Tikhonov}}{1994}]{1994A&A...286..718K}
{Karachentsev} I.~D.,  {Tikhonov} N.~A.,  1994, \aap, 286, 718

\bibitem[\protect\citeauthoryear{{Kawakita}, {Kinugasa}, {Ayani} \&
  {Yamaoka}}{{Kawakita} et~al.}{2004}]{2004IAUC.8266....2K}
{Kawakita} H.,  {Kinugasa} K.,  {Ayani} K.,    {Yamaoka} H.,  2004, \iauc,
  8266, 2

\bibitem[Leonard et al.(2003)]{2003ApJ...594..247L} Leonard, D.~C., Kanbur, 
S.~M., Ngeow, C.~C., \& Tanvir, N.~R.\ 2003, \apj, 594, 247 

\bibitem[\protect\citeauthoryear{{Leonard}, {Filippenko}, {Gates}, {Li},
  {Eastman}, {Barth}, {Bus}, {Chornock} \& {et al.}}{{Leonard}
  et~al.}{2002}]{2002PASP..114...35L}
{Leonard} D.~C.,  {Filippenko} A.~V.,  {Gates} E.~L.,  {Li} W.,  {Eastman}
  R.~G.,  {Barth} A.~J.,  {Bus} S.~J.,  {Chornock} R.,    {et al.} 2002, \pasp,
  114, 35

\bibitem[\protect\citeauthoryear{{Leonard}, {Filippenko}, {Li}, {Matheson},
  {Kirshner}, {Chornock}, {Van Dyk}, {Berlind} \& {et al.}}{{Leonard}
  et~al.}{2002}]{2002AJ....124.2490L}
{Leonard} D.~C.,  {Filippenko} A.~V.,  {Li} W.,  {Matheson} T.,  {Kirshner}
  R.~P.,  {Chornock} R.,  {Van Dyk} S.~D.,  {Berlind} P.,    {et al.} 2002,
  \aj, 124, 2490


\bibitem[\protect\citeauthoryear{{Li}, {Van Dyk}, {Filippenko}, {Cuillandre},
  {Jha}, {Bloom}, {Riess} \& {Livio}}{{Li} et~al.}{2005a}]{2005astro.ph..7394L}
{Li} W.,  {Van Dyk} S.~D.,  {Filippenko} A.~V.,  {Cuillandre} J.,  {Jha} S.,
  {Bloom} J.~S.,  {Riess} A.~G.,    {Livio} M.,  2005c, ArXiv Astrophysics
  e-prints 0507394

\bibitem[\protect\citeauthoryear{{Li}, {Van Dyk}, {Filippenko} \&
  {Cuillandre}}{{Li} et~al.}{2005b}]{2005PASP..117..121L}
{Li} W.,  {Van Dyk} S.~D.,  {Filippenko} A.~V.,    {Cuillandre} J.,  2005a,
  \pasp, 117, 121

\bibitem[\protect\citeauthoryear{{Litvinova} \& {Nadyozhin}}{{Litvinova} \&
  {Nadyozhin}}{1985}]{1985SvAL...11..145L}
{Litvinova} I.~Y.,  {Nadyozhin} D.~K.,  1985, Soviet Astronomy Letters, 11, 145

\bibitem[Ma{\'{\i}}z-Apell{\'a}niz et al.(2004)]{2004ApJ...615L.113M} 
Ma{\'{\i}}z-Apell{\'a}niz, J., Bond, H.~E., Siegel, M.~H., Lipkin, Y., 
Maoz, D., Ofek, E.~O., \& Poznanski, D.\ 2004, \apjl, 615, L113 

\bibitem[\protect\citeauthoryear{{Mathis}}{{Mathis}}{2000}]{2000asqu.book.....%
M}
{Mathis} J.~S.,  2000, {in Allen's Astrophysical Quantities}.
4th ed. Springer, Edited by Arthur N.~Cox, p523

\bibitem[\protect\citeauthoryear{Maund}{{Maund}}{2005}]{jrmthesis}
{Maund} J.~R., 2005, PhD Thesis, University of Cambridge

\bibitem[\protect\citeauthoryear{{Maund} \& {Smartt}}{{Maund} \&
  {Smartt}}{2005}]{2005MNRAS.360..288M}
{Maund} J.~R.,  {Smartt} S.~J.,  2005, \mnras, 360, 288

\bibitem[Maund et al.(2005)]{2005MNRAS.tmpL..88M} Maund, J.~R., Smartt, 
S.~J., \& Danziger, I.~J.\ 2005, \mnras, L88 

\bibitem[\protect\citeauthoryear{{Maund}, {Smartt}, {Kudritzki},
  {Podsiadlowski} \& {Gilmore}}{{Maund} et~al.}{2004}]{2004Natur.427..129M}
{Maund} J.~R.,  {Smartt} S.~J.,  {Kudritzki} R.~P.,  {Podsiadlowski} P.,
  {Gilmore} G.~F.,  2004, \nat, 427, 129

\bibitem[\protect\citeauthoryear{{Meynet}, {Maeder}, {Schaller}, {Schaerer} \&
  {Charbonnel}}{{Meynet} et~al.}{1994}]{1994A&AS..103...97M}
{Meynet} G.,  {Maeder} A.,  {Schaller} G.,  {Schaerer} D.,    {Charbonnel} C.,
  1994, \aaps, 103, 97

\bibitem[\protect\citeauthoryear{{Nadyozhin}}{{Nadyozhin}}{2003}]{2003MNRAS.34%
6...97N}
{Nadyozhin} D.~K.,  2003, \mnras, 346, 97

\bibitem[\protect\citeauthoryear{{Nakano}, {Itagaki}, {Kushida} \&
  {Kushida}}{{Nakano} et~al.}{2004}]{2004IAUC.8265....1N}
{Nakano} S.,  {Itagaki} K.,  {Kushida} R.,    {Kushida} Y.,  2004, \iauc, 8265,
  1

\bibitem[\protect\citeauthoryear{{Pastorello}, {Ramina}, {Zampieri},
  {Navasardyan}, {Salvo} \& {Fiaschi}}{{Pastorello}
  et~al.}{2004a}]{astro-ph/0310056}
{Pastorello} A.,  {Ramina} M.,  {Zampieri} L.,  {Navasardyan} H.,  {Salvo} M.,
    {Fiaschi} M.,  2004a, in Cosmic Explosions. On the 10th Anniversary of
  SN~1993J (IAU Colloquium 192). Springer, Heidelberg, p. 195 {Observational
  Properties of Type II Plateau Supernovae, astro-ph/0310056}

\bibitem[\protect\citeauthoryear{{Pastorello}, {Zampieri}, {Turatto},
  {Cappellaro}, {Meikle}, {Benetti}, {Branch}, {Baron}, {Patat}, {Armstrong},
  {Altavilla}, {Salvo} \& {Riello}}{{Pastorello}
  et~al.}{2004b}]{2004MNRAS.347...74P}
{Pastorello} A.,  {Zampieri} L.,  {Turatto} M.,  {Cappellaro} E.,  {Meikle}
  W.~P.~S.,  {Benetti} S.,  {Branch} D.,  {Baron} E.,  {Patat} F.,  {Armstrong}
  M.,  {Altavilla} G.,  {Salvo} M.,    {Riello} M.,  2004b, \mnras, 347, 74

\bibitem[Pilyugin et al.(2004)]{2004A&A...425..849P} Pilyugin, L.~S., 
V{\'{\i}}lchez, J.~M., \& Contini, T.\ 2004, \aap, 425, 849 

\bibitem[Rajala et al.(2005)]{2005PASP..117..132R} Rajala, A.~M., et al.\ 
2005, \pasp, 117, 132 

\bibitem[\protect\citeauthoryear{{Rozanski} \& {Rowan-Robinson}}{{Rozanski} \&
  {Rowan-Robinson}}{1994}]{1994MNRAS.271..530R}
{Rozanski} R.,  {Rowan-Robinson} M.,  1994, \mnras, 271, 530

\bibitem[\protect\citeauthoryear{{Schaller}, {Schaerer}, {Meynet} \&
  {Maeder}}{{Schaller} et~al.}{1992}]{1992A&AS...96..269S}
{Schaller} G.,  {Schaerer} D.,  {Meynet} G.,    {Maeder} A.,  1992, \aaps, 96,
  269

\bibitem[\protect\citeauthoryear{{Schlegel}, {Finkbeiner} \&
  {Davis}}{{Schlegel} et~al.}{1998}]{1998ApJ...500..525S}
{Schlegel} D.~J.,  {Finkbeiner} D.~P.,    {Davis} M.,  1998, \apj, 500, 525

\bibitem[\protect\citeauthoryear{{Smartt}, {Gilmore}, {Tout} \&
  {Hodgkin}}{{Smartt} et~al.}{2002}]{2002ApJ...565.1089S}
{Smartt} S.~J.,  {Gilmore} G.~F.,  {Tout} C.~A.,    {Hodgkin} S.~T.,  2002,
  \apj, 565, 1089

\bibitem[\protect\citeauthoryear{{Smartt}, {Gilmore}, {Trentham}, {Tout} \&
  {Frayn}}{{Smartt} et~al.}{2001}]{2001ApJ...556L..29S}
{Smartt} S.~J.,  {Gilmore} G.~F.,  {Trentham} N.,  {Tout} C.~A.,    {Frayn}
  C.~M.,  2001, \apjl, 556, L29

\bibitem[\protect\citeauthoryear{{Smartt}, {Maund}, {Gilmore}, {Tout},
  {Kilkenny} \& {Benetti}}{{Smartt} et~al.}{2003}]{2003MNRAS.343..735S}
{Smartt} S.~J.,  {Maund} J.~R.,  {Gilmore} G.~F.,  {Tout} C.~A.,  {Kilkenny}
  D.,    {Benetti} S.,  2003, \mnras, 343, 735

\bibitem[\protect\citeauthoryear{{Smartt}, {Maund}, {Hendry}, {Tout},
  {Gilmore}, {Mattila} \& {Benn}}{{Smartt} et~al.}{2004}]{2004Sci...303..499S}
{Smartt} S.~J.,  {Maund} J.~R.,  {Hendry} M.~A.,  {Tout} C.~A.,  {Gilmore}
  G.~F.,  {Mattila} S.,    {Benn} C.~R.,  2004, Science, 303, 499

\bibitem[\protect\citeauthoryear{{Sohn} \& {Davidge}}{{Sohn} \&
  {Davidge}}{1996}]{1996AJ....111.2280S}
{Sohn} Y.,  {Davidge} T.~J.,  1996, \aj, 111, 2280

\bibitem[\protect\citeauthoryear{{Stetson}}{{Stetson}}{1987}]{1987PASP...99..191S}
{Stetson} P.~B., 1987, \pasp, 99, 191


\bibitem[\protect\citeauthoryear{{Stetson} \& {Harris}}{{Stetson} \&
  {Harris}}{1988}]{1988AJ.....96..909S}
{Stetson} P.~B.,  {Harris} W.~E.,  1988, \aj, 96, 909

\bibitem[\protect\citeauthoryear{{Theureau}, {Bottinelli}, {Coudreau-Durand},
  {Gouguenheim}, {Hallet}, {Loulergue}, {Paturel} \& {Teerikorpi}}{{Theureau}
  et~al.}{1998}]{1998A&AS..130..333T}
{Theureau} G.,  {Bottinelli} L.,  {Coudreau-Durand} N.,  {Gouguenheim} L.,
  {Hallet} N.,  {Loulergue} M.,  {Paturel} G.,    {Teerikorpi} P.,  1998,
  \aaps, 130, 333

\bibitem[\protect\citeauthoryear{{Thielemann}, {Brachwitz}, {H{\"o}flich},
  {Martinez-Pinedo} \& {Nomoto}}{{Thielemann}
  et~al.}{2004}]{2004NewAR..48..605T}
{Thielemann} F.-K.,  {Brachwitz} F.,  {H{\"o}flich} P.,  {Martinez-Pinedo} G.,
    {Nomoto} K.,  2004, New Astronomy Review, 48, 605

\bibitem[\protect\citeauthoryear{{Tonry}, {Blakeslee}, {Ajhar} \&
  {Dressler}}{{Tonry} et~al.}{2000}]{2000ApJ...530..625T}
{Tonry} J.~L.,  {Blakeslee} J.~P.,  {Ajhar} E.~A.,    {Dressler} A.,  2000,
  \apj, 530, 625

\bibitem[\protect\citeauthoryear{{Tully}}{{Tully}}{1988}]{1988ngc..book.....T}
{Tully} R.~B.,  1988, {Nearby galaxies catalog}.
Cambridge and New York, Cambridge University Press, 1988, 221 p.

\bibitem[\protect\citeauthoryear{{Tully} \& {Shaya}}{{Tully} \&
  {Shaya}}{1984}]{1984ApJ...281...31T}
{Tully} R.~B.,  {Shaya} E.~J.,  1984, \apj, 281, 31

\bibitem[\protect\citeauthoryear{{Turatto}, {Mazzali}, {Young}, {Nomoto},
  {Iwamoto}, {Benetti}, {Cappallaro}, {Danziger} \& {et al.}}{{Turatto}
  et~al.}{1998}]{1998ApJ...498L.129T}
{Turatto} M.,  {Mazzali} P.~A.,  {Young} T.~R.,  {Nomoto} K.,  {Iwamoto} K.,
  {Benetti} S.,  {Cappallaro} E.,  {Danziger} I.~J.,    {et al.} 1998, \apjl,
  498, L129+

\bibitem[Urbaneja et al.(2005)]{Urbaneja_m33_gradient} Urbaneja, M.~A., et 
al.\ 2005, \apj, in press 

\bibitem[Van Dyk et al.(2002)]{2002PASP..114.1322V} Van Dyk, S.~D., 
Garnavich, P.~M., Filippenko, A.~V., H{\"o}flich, P., Kirshner, R.~P., 
Kurucz, R.~L., \& Challis, P.\ 2002, \pasp, 114, 1322 

\bibitem[Van Dyk et al.(2003)]{2003PASP..115....1V} Van Dyk, S.~D., Li, W., 
\& Filippenko, A.~V.\ 2003a, \pasp, 115, 1 

\bibitem[Van Dyk et al.(2003)]{2003PASP..115.1289V} Van Dyk, S.~D., Li, W., 
\& Filippenko, A.~V.\ 2003b, \pasp, 115, 1289 

 \bibitem[Vila-Costas \& Edmunds(1992)]{1992MNRAS.259..121V} Vila-Costas, 
M.~B., \& Edmunds, M.~G.\ 1992, \mnras, 259, 121 

\bibitem[\protect\citeauthoryear{{Walborn}, {Prevot}, {Prevot}, {Wamsteker},
  {Gonzalez}, {Gilmozzi} \& {Fitzpatrick}}{{Walborn}
  et~al.}{1989}]{1989A&A...219..229W}
{Walborn} N.~R.,  {Prevot} M.~L.,  {Prevot} L.,  {Wamsteker} W.,  {Gonzalez}
  R.,  {Gilmozzi} R.,    {Fitzpatrick} E.~L.,  1989, \aap, 219, 229

\bibitem[\protect\citeauthoryear{{White} \& {Malin}}{{White} \&
  {Malin}}{1987}]{1987Natur.327...36W}
{White} G.~L.,  {Malin} D.~F.,  1987, \nat, 327, 36

\bibitem[\protect\citeauthoryear{{Zampieri}, {Pastorello}, {Turatto},
  {Cappellaro}, {Benetti}, {Altavilla}, {Mazzali} \& {Hamuy}}{{Zampieri}
  et~al.}{2003}]{2003MNRAS.338..711Z}
{Zampieri} L.,  {Pastorello} A.,  {Turatto} M.,  {Cappellaro} E.,  {Benetti}
  S.,  {Altavilla} G.,  {Mazzali} P.,    {Hamuy} M.,  2003, \mnras, 338, 711

\bibitem[\protect\citeauthoryear{{Zampieri}, {Ramina} \&
  {Pastorello}}{{Zampieri} et~al.}{2004}]{astro-ph/0310057}
{Zampieri} L.,  {Ramina} M.,    {Pastorello} A.,  2004, in Supernovae (10 Years
  of 1993J), Marcaide J. M., Weiler K. W., eds, Proc. IAU Coll. 192,
  Springer-Verlag, Berlin, in press, {Understanding Type II Supernovae,
  astro-ph/0310057}

\end{thebibliography}
\end{document}